\documentclass[superscriptaddress,onecolumn]{revtex4}
\usepackage{mathrsfs}
\usepackage{amssymb}
\usepackage[tbtags]{amsmath}
\usepackage{CJK}
\usepackage{color}
\usepackage{mathtools}
\usepackage{subfigure}
\begin{document}
\title{Simulating Dirac equation with Josephson junction circuits}
\author{Xiao hui Ji}
\email{xiaohui.cc.ji@gmail.com}
\affiliation{School of Materials and Engineering, Southwest jiaotong University Chengdu 610031, PR China}
\affiliation{Key Laboratory of Advanced Technology for Materials of Education Ministry, Southwest jiaotong University Chengdu 610031, PR China}
\author{Wen bin Lin}
\affiliation{School of Physical Science and Technology, Southwest Jiaotong University, Chengdu 610031, China}
\author{Jia gang Zeng}
\affiliation{School of Physical Science and Technology, Southwest Jiaotong University, Chengdu 610031, China}
\author{Guang di Wang}
\affiliation{School of Civil Engineering, Southwest Jiaotong University, Chengdu 610031, China}
\date{\today}
\begin{abstract}
We propose a scheme for simulating 3+1, 2+1, 1+1 Dirac equation for a free spin-1/2 particle with superconducting josephson circuits consisting of five qubits, four qubits, two qubits respectively. In 3+1D and 2+1D, the flux qubit1 driven by a resonant pulse is in the superposition state of its own two eigenstatesis, and it is used as a bus to induce the (blue)red-sideband excitation consisting of a magnetic pulse acting resonantly on two levels of the flux qubit2 and the energy levels of one phase qubit, which yields two (Anti)Jaynes-Cummings interactions with one driving pulse and reduces the damage of the driving pulses to the system consequently. Numerical results show that decoherence time is several times longer than transition time supposing set appropriate experimental parameters. Therefore experiments verifying the dynamics of electron and neutrino, such as {\em Zitterberwung} effect in 3+1, 2+1 and 1+1 dimensions, can be implemented by microelectronic chips composed of the qubits as artificial atoms.
\\
\end{abstract}
\maketitle
\par Due to the importance of Dirac equations, the simulations of Dirac equation are attracting more attentions.
The pioneering schemes for the 1+1 and 3+1 Dirac equation in trap ion system are proposed in 2007~\cite{1,2}. The 1+1 Dirac equation has been experimentally implemented in trapped ions system and the corresponding {\em Zitterberwung} effect has been observed~\cite{3}. The propagations of photon in 2D photonic crystals and acoustic wave in the 2D sonic crystals obey the 2+1 Dirac equation for m=0 and spin-1 and the Dirac tremor has been found to occur at the Dirac point of the crystals~\cite{4,5}. However, there is no experimental simulation for the 3+1 Dirac equation for a spin-1/2 particle so far.
\par Inspired by the original ideas that Josephson junctions circuits simulate natural atom~\cite{10,11} and the trapped ions simulate the Dirac equations~\cite{1,2,3,12,13}, in this {\em Letter} we propose an convenient scheme for simulating the 3+1, 2+1, and 1+1 Dirac equation based on the superconducting Josephson junctions circuits. Josephson devices can behave like artificial atoms~\cite{10,11}, meanwhile possess new features~\cite{15}. Under the present low temperature and the fabrication technology, Josephson devices have been used widely in the fields of quantum information~\cite{16} and quantum simulation~\cite{17,19}.
\par The proposed scheme is based on the unified description of {\em Zitterberwung} effect~\cite{13}.
The circuit diagram is shown in Fig.\,1. The 3+1D, 2+1D and 1+1D simulations can be implemented with the structures of
five qubits, four qubits, two qubits respectively. The levels of magnetic flux qubits are
mapped to the positive and negative energy levels, and the two spin levels of the particle.
The particle momentum in each dimension corresponds to a phase qubit, and the momentum operator
in every dimension is proportional to the translational operator constructed by the phase of its corresponding phase qubit.
Fig.\,2 presents the simulation results for the observations and the suggested parameters of the circuits.
\begin{figure}[htbp]
\centering
\includegraphics[width=3.5in,angle=-90]{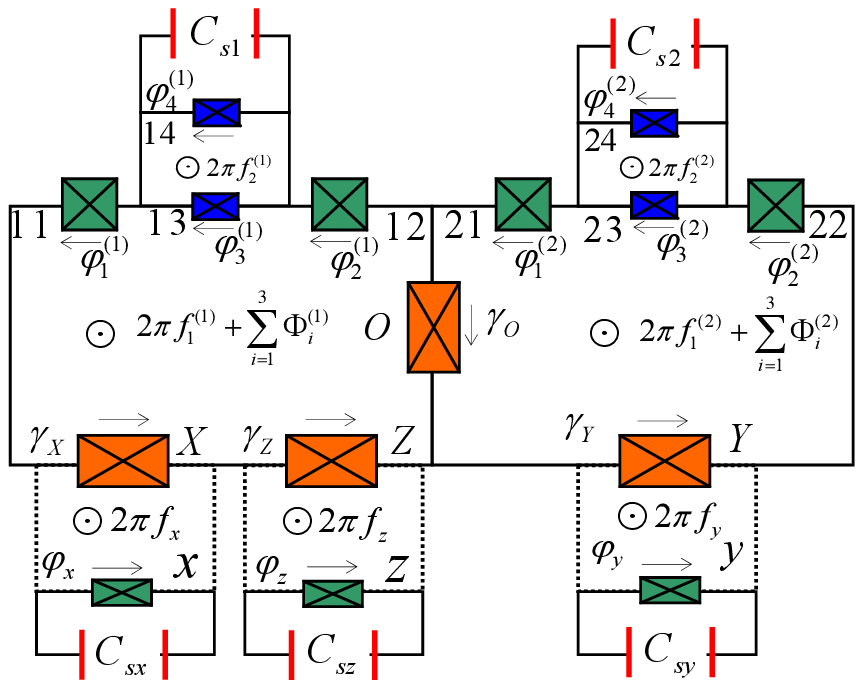}
\caption{(color online) Electrical schematic of the 3D Dirac equation simulations.
E-mail icon ($\times$ in $\square$) represents the Josephson junctions (JJ).
The black solid and dotted lines represent the superconductive wires.
The $l$th flux qubit consists of JJ $l1$ (green), $l2$ (green), $l3$ (blue),
 $l4$ (blue) and the superconductivity wires with inductance $L_g^{(l)}$.
 Here $l=1~\text{or}~2$. Both JJ $l1$ and $l2$ have Josephson energy $E_J^{(l)}$
 and capacitance $C_J^{(l)}$, and JJ $l3$ and $l4$ are smaller than $l1$ (or $l2$)
 by a factor ${\alpha _l}$. The loop enclosing JJ $l1$, $l2$ and $l3$ is applied with
 the reduced flux $2\pi f_1^{(l)}$ and the weak time-dependent magnetic fluxes $\Phi _i^{(l)}$(driving pulse),
 where $\Phi _i^{(l)} = n_i^{(l)}\cos (\omega _i^{(l)}t + \phi _i^{(l)})$, $i=1,2,3$.
${\varphi_i^{(l)}}$ denotes the phase differences across $i$th junction in $l$th qubit loop
and ${\varphi_L^{(l)}}$ is the phase differences due to the inductance of superconductivity ring
in $l$th qubit loop. The SQUID loop, consisted of JJ $l3$ and $l4$ which are parallel to
the capacitor $C_{sl}$ (red), is biased by the reduced flux $2\pi f_2^{(l)}$.
The shunt capacitance $C_{sl}$ is $({\beta_l} - 2{\alpha_l})$ times of Josephson
junction capacitance. The two flux qubits couple to each other with the shared
large Josephson junction $O$ (orange) which has Josephson energy $E_{JO}$ and
capacitance $C_{JO}$. The phase qubit $p$, parallel to a single crystal silicon
capacitor $C_{sp}$(red), has Josephson energy $E_{Jp}$ and capacitance $C_{Jp}$,
and the shared large junction $P$ has Josephson energy $E_{JP}$ and capacitance $C_{JP}$.
Here $p=x,~y,~z$ and $P=X,~Y,~Z$. The josephson energies satisfy ${E_{JP}},\; {E_{JO}} \gg {E_{J}^{(l)}} ,\;{E_{Jp}}$.
The loop, containing the phase qubit $p$, the shared large junction $P$ and the superconductivity wire
(the dotted line) with inductance $L_p$, is applied with the reduced flux
$2\pi f_p$, which provides a current offset for the phase qubit $p$.
$\varphi_p$, $\gamma _O$, $\gamma _P$ denote the phase differences across the phase qubit $p$,
shared junction $P$, shared junction $O$ and $\varphi_{Lp}$ is the phase differences due to the inductance $L_{p}$.
All of the shared junctions in the scheme can be replaced by the superconducting wires with
the same inductive values. \emph{} The first flux qubit ($l=1$) interacts with the phase qubits
$p=x,~z$ (green) with the shared large Josephson junctions $P=X,~Z$ (orange), and the second flux
qubit ($l=2$) interacts with the phase qubit $p=y$ (green) with the shared large Josephson junction $P=Y$ (orange).
This 3D electrical schematic reduces to the 2D electrical schematic by removing the loop
containing the phase qubit $z$, the shared large junction $Z$, and further reduces to the
1D case by removing the second flux qubit ($l=2$) and the shared large Josephson junction $O$ (orange).
}
\label{*}
\end{figure}
\begin{figure}[tbp]
\includegraphics[width=2in,height=3.5in,angle=-90]{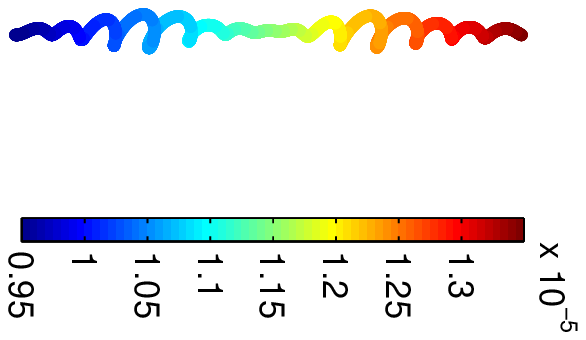}\\
\includegraphics[width=2in,height=3.5in,angle=-90]{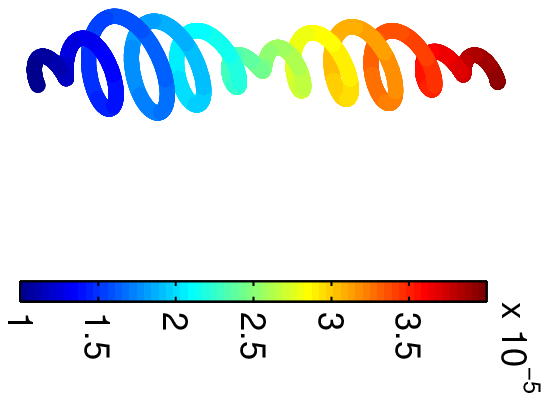}\\
\includegraphics[width=3.5in,height=2in]{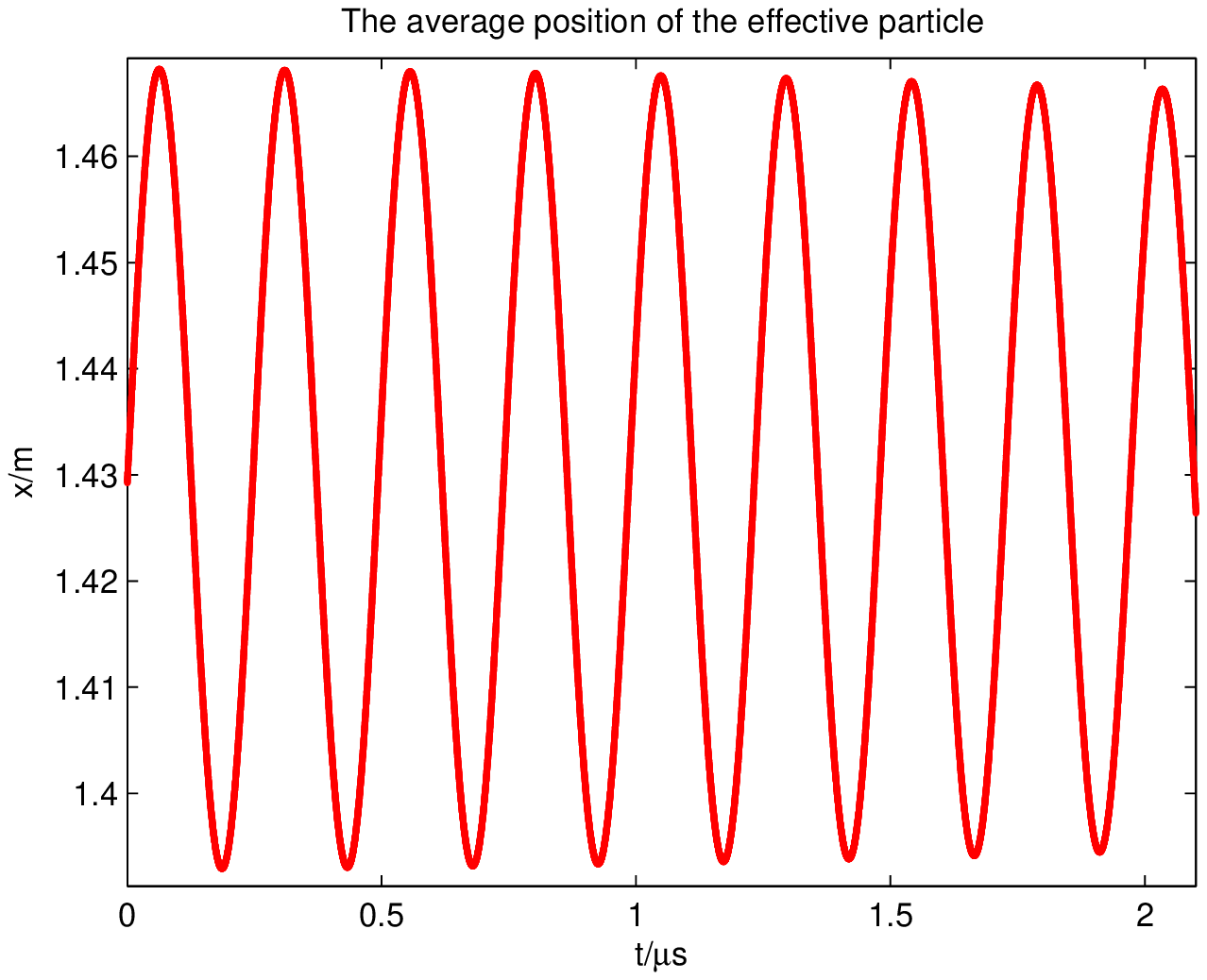}
\caption{The results similar to that shown in figure can be obtained by observing the circuits in Fig 1.
In the figure, the average position of ion with spin $1/2$, rest mass $m\ne 0$ and
$m=0$ are given respectively. The top graph shows $m\ne 0$ particles'
3+1 dimensional average position, the middle graph represents $m=0$ particles'
3+1 dimensional average position, the bottom graph depicts $m \ne 0$ particles' 1+1 dimensional average position.
$E_J^{(1)}/h=300GHz$, $E_J^{(2)}/h=400GHz$, $E_C^{(1)}=e^2/(2C_J^{(1)})$, $E_C^{(2)}=e^2/(2C_J^{(2)})$,
$E_J^{(1)}/E_C^{(1)}=30$, $E_J^{(2)}/E_C^{(2)}=30$, $\beta _1\sim 6$, $\beta _2\sim 6$,
$\alpha _1=0.6$, $\alpha _2=0.6$, $f_1^{(1)}=f_1^{(2)}\in[1/3+\Delta f/2-0.2,1/3+\Delta f/2+0.2]$,
$f_2^{(1)}=f_2^{(2)}=1/3-\Delta f$, $\Delta f=0.015$,
 The stable region: qubit1 $f_1^{(1)}+\Delta f/2\pm 0.02$; quibt2 $f_1^{(2)}+\Delta f/2\pm 0.2$.
 geometric inductance $L_g^{(1)}\sim33 pH$, $L_g^{(2)}\sim33 pH$, ${\Lambda _1} = {L_r^{(1)}}/{L_J^{(1)}}\sim0.17$,
 ${\Lambda _2} = {L_r^{(2)}}/{L_J^{(2)}}\sim0.23$,
 $L_r^{(l)} = L_g^{(l)}+\sum\nolimits_P{{L_{JP}}}+{L_{JO}}$. Where,
$l=1$ and $P=X,Z$, $l=2$ and $P=Y$ in 3+1 dimension;
$l=1$ and $P=X$, $l=2$ and $P=Y$ in 2+1 dimension;
$L_{JO}=0$ and $P=X$ in 1+1 dimension. $L_{Jp}$, $L_{JP}$, $L_{JO}$ denote effective
inductance of phase qubit $p$, shared junction $P$, shared junction $O$ respectively.
The phase qubit parameter: $p=x, y, z$, $P=X, Y, Z$; ${E_{Jp}}/h=850 \sim 1350GHz$,
$f_p$ is set properly so that $\sin {\varphi _{p0}}=I_{pb}/I_{p0}=0.99$,
 $I_{pb}$ bias current, $I_{p0}$ critical current. $E_{Jp}/E_{Cp}=10^{-6}$, $E_{Cp}=e^2/(C_{Jp}+C_{sp})$,
 ${E_{Lp}} = {({\Phi _0}/2\pi )^2}/(2{L_p})$, the total inductance of the ring enclosing the phase qubit $p$
 and shared junction $P$ is $L_{rp}=L_p+L_{JP}$, geometric inductance $L_p\sim40 pH$.
All shared junction are biased by no current source.
The shared junction parameter: ${E_{JP}}/h \sim 8000 GHz$, $\varphi _{P0}=\arcsin {(I_{Pb}/I_{P0})}$,
$I_{Pb}$ bias current, $I_{P0}$ critical current. ${E_{JO}}/h \sim 8000 GHz$,
$\varphi _{O0}=\arcsin {(I_{Ob}/I_{O0})}$, $I_{Ob}$ bias current, $I_{O0}$ critical current.
$n_x, n_y, n_z\sim0.005-0.025$, weak radio-frequency field $\Phi _i^{(l)}\sim0.005-0.025\Phi_0$.
$mc^2$ the rest energy of ion , $cp$ the kinetic energy.
3+1D and 2+1D: $mc^2=0\sim 5MHz$, $cp=0.02\sim0.1MHz$;
1+1D: $mc^2=1\sim 5MHz$, $cp=0.01\sim 10MHz$
}
\label{*}
\end{figure}
\par Here, we focus on the mutual coupling caused by the shared junction and
the self-coupling is written directly. 
The phase constraint conditions in the $l$th flux qubit loop and in $p$th phase qubit loop give
\begin{small}
\[\begin{array}{l}
2\pi f_1^{\scriptscriptstyle{(l)}}+\varphi _1^{\scriptscriptstyle{(l)}}+
\varphi _2^{\scriptscriptstyle{(l)}}+\varphi _3^{\scriptscriptstyle{(l)}}+\varphi _r^{\scriptscriptstyle{(l)}} +
 \sum\nolimits_{i ={r_l}}\!\!{\Phi _i^{\scriptscriptstyle{(l)}}}=0\\
\varphi _r^{\scriptscriptstyle{(l)}}\!\!=\!\!\varphi _L^{\scriptscriptstyle{(l)}}\!+\!{( - 1)^l}{\gamma _O}\!+\!
\sum\nolimits_{P}\!{{\gamma _P}},\;\;
\varphi _4^{\scriptscriptstyle{(l)}}\!\!-\!\!\varphi _3^{\scriptscriptstyle{(l)}}\!\!+\!\!2\pi f_2^{\scriptscriptstyle{(l)}}\!\!=\!\!0\\
{\varphi _p}+\varphi_{Lp}-{\gamma _P} + 2\pi {f_p}=0
\end{array}\]
\end{small}
where lowercase $p$ corresponds to uppercase $P$;
3+1D:
\begin{small}$r_l\!\!=\!\!1,2,3;\;l\!\!=\!\!1\!\!:\!\!P\!\!=\!\!X,Z;\;l\!\!=\!\!2\!\!:\!\!P\!\!=\!\!Y.$\end{small}
\;\;2+1D:
\begin{small}$r_l\!=\!1,2;\;l\!=\!1\!\!:\!\!P\!=\!X;\;l\!=\!2\!\!:\!\!P\!=\!Y.$\end{small}
\;\;1+1D:
\begin{small}$r_l\!=\!1,2;\;\gamma _O\!=\!0;\;l\!=\!1\!\!:\!\!P\!=\!X.$\end{small}
To be simplified, the dynamics of the models representing 3+1, 2+1, 1+1 Dirac
equations can be generally written as the Hamiltonian
\begin{small}
\begin{equation}
H = \sum\limits_{l = L} {H_q^{(l)}}  + \sum\limits_{p = c} {{H_p}}
+ \sum\limits_{P = C} {{H_P}} +\sum\limits_{pP} {{H_{pP}}} + {H_O}
\end{equation}
\end{small}
\begin{small}
\[\begin{array}{l}
3\!+\! 1D:L = 2;C = X,Y,Z;c = x,y,z;\\
\;\;\;\;\;\;\;\;\;\;\;\;\;pP:p = x\& P = X;p = y\& P = Y;p = z\& P = Z;\\
2\!+\!1D:L = 2;C = X,Y,T;c = x,y,z;\\
\;\;\;\;\;\;\;\;\;\;\;\;\;pP:p = x\& P = X;p = y\& P = Y;\\
1\!+\!1D:L = 1;C = X;c = x;\;{H_O} = 0;\;pP\!:\!p= x\& P = X.
\end{array}\]
\end{small}
Here, $H_q^{(l)}$ in equation(1) is the Hamiltonian of single flux qubit~\cite{20,21,23,25,26,27}, and
it is shown as
\begin{small}
\[\begin{array}{l}
H_q^{(l)} \!\!=\!2E_{Ca}^{(l)}N_{al}^2 + 2E_{Cs}^{(l)}N_{sl}^2 - 2E_J^{(l)}\cos \varphi _a^{(l)}\cos \varphi _s^{(l)}\\
\;\;\;\;\;\; - 2{\alpha _1}E_J^{(l)}\!\cos (\pi f_2^{(l)})\!\cos (2\varphi _s^{(l)}\!\!+\!\!
2\pi f_3^{(l)}\!\!+\!\!\varphi _r^{(l)} \!\!+\!\!\sum\limits_{i = 1}^3 {\Phi _i^{(l)}} )\\
\;\;\;\;\;\; + [2{\beta _1}/(1 + 4{\beta _l})]\hbar N_s^{(l)}\!\!\sum\limits_{i = 1}^3 {\dot \Phi _i^{(l)}}  +
 \frac{1}{2}L_r^{(l)}{[{I^{(l)}}]^2}
\end{array}\]
\end{small}
The circling current is shown as~\cite{20,21,23,25,26,27,29,31}
\begin{small}
\[{I^{(l)}} = (2e)({{\dot N}_{sl}} + {{\dot N}_{al}}) + (2\pi /{\Phi _0})E_J^{(l)}\sin (\varphi _s^{(l)} + \varphi _a^{(l)})\]
\end{small}
charging energy
\begin{small}$E_{Ca}^{(l)} = {e^2}/2C_J^{(l)}$\end{small},
\begin{small}$E_{Cs}^{(l)} = {e^2}/[2{C_J^{(l)}}(1 + 4{\beta _l})]$\end{small},
cooper pair operator
\begin{small}${N_{al}} =  - i\partial /\partial \varphi _a^{(l)}$\end{small},
\begin{small}${N_{sl}} =  - i\partial /\partial \varphi _s^{(l)}$\end{small},
new phase variables
\begin{small}$\varphi_a^{(l)}\!\!=\!\!(\varphi _1^{(l)}\!\!-\!\!\varphi_2^{(l)})/2,\;
\varphi_s^{(l)}\!\!=\!\!(\varphi_1^{(l)}\!\!+\!\!\varphi_2^{(l)})/2$\end{small},
flux bias \begin{small}$f_3^{(l)}= f_1^{(l)}+f_2^{(l)}/2$\end{small},
\begin{small}${\Phi _0}$\end{small} magnetic flux quantum.
\begin{small}$\hbar=h/2\pi$\end{small}, \begin{small}$h$\end{small} is planck constant.
$e$ is electronic charge.
Here, $H_p$ in equation(1) is the Hamiltonian of $p$th phase qubit, and it is described as \cite{29}
\begin{small}
\[{H_p}\!=\!4{E_{cp}}N_p^2\!-\!{E_{Jp}}(\cos {\varphi _p}\!+\!{I_{pb}}{\varphi _p}/{I_{p0}})
\!+\!{E_{rp}}{({\varphi _p}\!-\!{\varphi _{p0}})^2}/2\]
\end{small}
charging energy \begin{small}${E_{cp}} = {e^2}/2({C_p} + {C_{sp}})$\end{small},
cooper pair operator \begin{small}${N_{p}}\!\!=\!-\!i\partial/\partial\!{\varphi _{p}}$\end{small},
the coupling energy \begin{small}${E_{rp}}\!=\!{({\Phi _0}/2\pi )^2}/{L_{rp}}$\end{small}.
Here, $H_P$ in equation(1) is the Hamiltonian of $P$th shared phase junction, and it is described as\cite{25,29}
\begin{small}
\[{H_P} = {({\Phi _0}/2\pi )^2}{C_P}\dot \gamma _P^2/2 - {E_{JP}}(\cos {\gamma _P} + {I_{Pb}}{\gamma _P}/{I_{P0}})\]
\end{small}
Here, $H_{pP}$ in equation(1) is the Hamiltonian of superconducting ring corresponding $p$th phase qubit
and shared junction $P$, and it is described as \cite{29}
\begin{small}
\[\begin{array}{l}
{H_{pP}}= ({E_{Lp}/2}){({\varphi _p} - {\gamma _P}+ 2\pi {f _p})^2}
\end{array}\]
\end{small}
The effective coupling energy is \begin{small}${E_{Lp}} = {({\Phi _0}/2\pi )^2}/{L_p}$\end{small}.
Here, $H_O$ in equation(1) is the Hamiltonian of $O$th big phase junction, and it is described as\cite{25,29}
\begin{small}
\[{H_O} = {({\Phi _0}/2\pi )^2}{C_O}\dot \gamma _P^2/2 - {E_{JO}}\cos {\gamma _O}+ {I_{Ob}}{\gamma _P}/{I_{O0}})\]
\end{small}
\par $H_{\scriptscriptstyle{q0}}^{\scriptscriptstyle{(l)}}$ indicates
the Hamiltonian of the $l$th bare magnetic flux qubit biased by
the static magnetic field. \begin{small}$|e_l\rangle $\end{small}, \begin{small}$|g_l \rangle$\end{small}
represent two lower energy eigenstates of \begin{small}$H_{q0}^{(l)}$\end{small}.
For convenience, the basis spaces
\begin{small}$\{|g_1\rangle, |e_1\rangle\}$\end{small},
\begin{small}$\{|g_2\rangle, |e_2\rangle\}$\end{small}
are transformed respectively into
\begin{small}$\{|1D\rangle, |1U\rangle\}$\end{small},
\begin{small}$\{|20\rangle, |21\rangle\}$\end{small},
\begin{small}$H_{q0}^{(1)}$\end{small}, \begin{small}$H_{q0}^{(2)}$\end{small}
are transformed respectively into
\begin{small}
\[H_{D0}^{(1)} =\hbar X_0^{(1)}\hat \sigma _x^{(1)},\;
H_{D0}^{(2)} =\hbar Z_0^{(2)}\hat \sigma _z^{(2)}.\]
\end{small}
The Pauli operators are defined by
\begin{small}
\[{\hat \sigma _z}\!\!=\!\!\left(\!\!
{\begin{array}{*{20}{c}}
1&0\\
0&{ - 1}
\end{array}} \!\!\right)\;\;
{\hat \sigma _x}\!\!=\!\!\left(\!\!
{\begin{array}{*{20}{c}}
0&1\\
1&0
\end{array}} \!\!\right)\;\;
{\hat \sigma _ + }\!\!=\!\!\left(\!\!
{\begin{array}{*{20}{c}}
0&1\\
0&0
\end{array}} \!\!\right)\;\;
{\hat \sigma _ - }\!\!=\!\!\left(\!\!
{\begin{array}{*{20}{c}}
0&0\\
1&0
\end{array}} \!\!\right)\]
\end{small}
In population representation, the Hamiltonian $H_p$ and photon energy are shown as
\begin{small}\[{H_{p}}\!\!=\!\!\hbar {\omega _p}(\hat a_p^ + {{\hat a}_p}\!\!+\!\! 1/2),\;
\hbar {\omega _p}\!\!=\!\!{[8{E_{Cp}}({E_{Jp}}\cos {\varphi _{p0}}\!\!+\!\!{E_{rp}})]^{1/2}}\]\end{small}
The boson operator is shown as
\begin{small}\[
\hat a_p^ +\!=\!{\varphi _{p1}}/(2{\lambda _p})\!-\!i{\lambda _p}{N_p},\;
{\lambda _p}\!=\!{[2{E_{Cp}}/({E_{Jp}}\cos\!{\varphi _{p0}}\!+\!{E_{rp}})]^{1/4}}
\]\end{small}
\begin{small}${|n_p\rangle}$\end{small} represent eigenstates of $p$ phase qubit\cite{29}.
Similarly, the Hamiltonian of the shared junction
\begin{small}
\[H_P={\hbar {\omega _P}\hat a_P^ + {{\hat a}_P}},\;
H_O=\hbar {\omega _O}\hat a_O^ + {{\hat a}_O}\]
\end{small}
\par Rotating frame by reference
\begin{small}
\begin{equation}
H=\sum\limits_{l=L}{H_{D0}^{(l)}}+\sum\limits_{p=c}{{H_p}}+\sum\limits_{P=C}{{H_P}}+{H_O}
\end{equation}
\end{small}
satisfying the conditions of the frequencies
\begin{small}
\[\begin{array}{l}
 + 2X_0^{(1)} - \omega _3^{(1)} = 0,\;\;\phi _3^{(1)} = 0,\;n_3^{(1)} = {n_m}\\
 + {\omega _z} - \omega _3^{(2)} = 0,\;\phi _3^{(2)} =  - \pi /2,\;n_3^{(2)} = {n_z}\\
 + 2Z_0^{(2)} + {\omega _y} - \omega _1^{(1)} = 0,\;\phi _1^{(1)} = 0,\;n_1^{(1)} = {n_y}\\
 + 2Z_0^{(2)} - {\omega _y} - \omega _2^{(1)} = 0,\;\phi _2^{(1)} = \pi ,\;n_2^{(1)} = {n_y}\\
 + 2Z_0^{(2)} + {\omega _x} - \omega _1^{(2)} = 0,\;\phi _1^{(2)} =  - \pi /2,\;n_1^{(2)} = {n_x}\\
 + 2Z_0^{(2)} - {\omega _x} - \omega _2^{(2)} = 0,\;\phi _2^{(2)} =  + \pi /2,\;n_2^{(2)} = {n_x}\\
{\omega _O} \gg {\omega _X} \ne {\omega _Y} \ne {\omega _Z} \gg 2X_0^{(1)},\;2Z_0^{(2)},\;\omega _i^{(l)}
\end{array}\]
\end{small}
then the iterative calculations are carried out according to the following formula
\begin{small}
\begin{equation}
U\!=\!1\!+\!\sum\limits_{n = 1}^\infty  {{{\!\left( {\frac{{ - i}}{\hbar }} \right)}^n}
\!\int_{{t_0}}^t {d{t_1}\!\cdots\!\int_{{t_0}}^{{t_{n - 1}}}\!{{H_I}({t_1})\!
\cdots\!{H_I}({t_n})d{t_n}} } }
\end{equation}
\end{small}
Establish mapping relationship between two systems:
\begin{small}
\begin{equation}
\begin{array}{l}
c \sim 2{\Delta _p}{{\tilde \Omega }_p},\;\;{\Delta _p} = {\lambda _p},\;\;{{\hat p}_p} = i\hbar (a_p^ +  - a_p^ - )/2{\Delta _p}\\
c{{\hat p}_p}\!\sim\!\hbar {{\tilde \Omega }_p}\!=\!2\pi\!\times\!{\lambda _p}{n_p}\hbar X_2^{\scriptscriptstyle{(l)}}\hbar X_2^{\scriptscriptstyle{(l)}}{E_{Lp}}{({E_{JO}}{E_{JX}})^{\scriptscriptstyle{-1}}}\\
c{{\hat p}_0} \sim \hat \sigma _x^{(1)}\hat \sigma _z^{(1)}c{{\hat p}_0} = 2\pi  \times \hbar X_1^{(1)}\hbar Z_1^{(1)}E_{JO}^{ - 1}\\
m{c^2} \sim \hbar \Omega  = 2\pi  \times Z_1^{(1)}{n_m},\;\;(p = x,\;y,\;z).
\end{array}
\end{equation}
\end{small}
Then Hamiltonian $H_{eff}$ obtained by the iteration is implemented by the diagonal transformation
\begin{small}
${H_I} = {D^{ - 1}}({\vartheta}){H_{eff}}D({\vartheta})$
\end{small} with
\begin{small}
\[D({\vartheta}) = \left[ {\begin{array}{*{20}{c}}
{\cos ({\vartheta}/2)}&{ - i\sin ({\vartheta}/2)}\\
{i\sin ({\vartheta}/2)}&{ - \cos ({\vartheta}/2)}
\end{array}} \right],\]
\end{small}
one can obtain
\begin{small}
\[{H_I} = m{c^2}\hat \sigma _z^{(1)}{{\rm I}^{(2)}} + \hat \sigma _x^{(1)}
{{\hat \sigma }^{(2)}}cp + \hat \sigma _x^{(1)}\hat \sigma _z^{(1)}c{p_0}\]
\end{small}
The factors in the Hamiltonian are shown as
\begin{small}
\[\begin{array}{l}
{\theta} = \arccos (z_0^{(l)}{[{(z_0^{(l)})^2} + {(x_0^{(l)})^2}]^{ - 1/2}})\\
{\vartheta}\!=\!\arccos (Z_1^{(1)}{[{(Z_1^{(1)})^2}\!+\!{(Z_3^{(1)}{\omega _3})^2}]^{ - 1/2}})\\
Z_i^{(l)} = {m_l}zr_i^{(l)}\hat \sigma _z^{(l)},\;\;\;X_i^{(l)} = {m_l}xr_i^{(l)}\hat \sigma _x^{(l)}\\
when\;i = 0,1,2,\;{m_l} = {\mu _l};\;i = 3,\;{m_l} = {\nu _l};\;l=1,2\\
{\mu _l} = 2{\alpha _l}E_J^{(l)}\cos (\pi f_1^{(l)}),\;\;\;\;{\nu _l} = 2{\beta _l}/(1 + 4{\beta _l})\\
zr_i^{(1)}\!=\!x_i^{(1)}\!\cos\!{\theta _1}\!-\!z_i^{(1)}\!\sin\!{\theta _1},\;
xr_i^{(1)}\!\!=\!\!x_i^{(1)}\!\sin\!{\theta _1}\!+\! z_i^{(1)}\!\cos\!{\theta _1}\\
zr_i^{(2)}\!=\!z_i^{(2)}\!\cos\!{\theta _2}\!+\!x_i^{(2)}\!\sin\!{\theta _2},\;
xr_i^{(2)}\!\!=\!\!z_i^{(2)}\!\sin\!{\theta _2}\!-\! x_i^{(2)}\!\cos\!{\theta _2}\\
z_0^{(l)}=(\langle {e_l}|H_{q0}^{(l)}|{e_l}\rangle\!-\!\langle {g_l}|{H_{q0}^{(l)}}|{g_l}\rangle )/2,\;
x_0^{(l)}=\langle {e_l}|H_{q0}^{(l)}|{g_l}\rangle\\
z_1^{(l)} = (\langle {e_l}|si{n_l}|{e_l}\rangle-\langle {g_l}|si{n_l}|{g_l}\rangle )/2,\;x_1^{(l)} = \langle {e_l}|si{n_l}|{g_l}\rangle\\
z_2^{(l)} = (\langle {e_l}|co{s_l}|{e_l}\rangle-\langle {g_l}|co{s_l}|{g_l}\rangle )/2,\;x_2^{(l)} = \langle {e_l}|co{s_l}|{g_l}\rangle\\
{\sin _l} = sin(2\varphi _a^{(l)} + 2\pi f_3^{(l)})\;,\;{\cos _l} = \cos (2\varphi _a^{(l)} + 2\pi f_3^{(l)})\\
z_3^{(l)}\!=(\langle {e_l}|{N_{al}}|{e_l}\rangle\!-\!\langle {g_l}|{N_{al}}|{g_l}\rangle )\hbar/2,\;
x_3^{(l)}\!=\langle {e_l}|{\hbar N_{al}}|{g_l}\rangle
\end{array}\]
\end{small}
The basis space
\begin{small}$\{ \left| {1D} \right\rangle ,\left| {1U} \right\rangle \}$\end{small}
rotates to
\begin{small}$\{ \left| {10} \right\rangle ,\left| {11} \right\rangle \}$\end{small}.
After implementing Lorentz transformation S on the hamiltonian
\begin{small}
\begin{equation}
S = \left[ {\begin{array}{*{20}{c}}
{\cosh (az)}&0&{\sinh (az)}&0\\
0&{\cosh (az)}&0&{ - \sinh (az)}\\
{-\sinh (az)}&0&{\cosh (az)}&0\\
0&{\sinh (az)}&0&{\cosh (az)}
\end{array}} \right]
\end{equation}
\end{small}
with $az = 0.5\tanh^{-1} [ - (c{p_0}/m{c^2}){[1 + {(c{p_0}/m{c^2})^2}]^{ - 0.5}}]$,
the standard form can be obtained
\begin{small}
\begin{equation}
\begin{array}{l}
 H_I\!\!=\!\!\left[ {\begin{array}{*{20}{c}}
{m{c^2}}&0&{c{p_z}}&{c({p_x}\!\!-\!\!i{p_y})}\\
0&{m{c^2}}&{c({p_x}\!\!+\!\!i{p_y})}&{ - c{p_z}}\\
{c{p_z}}&{c({p_x}\!\!-\!\!i{p_y})}&{ - m{c^2}}&0\\
{c({p_x}\!\!+\!\! i{p_y})}&{ - c{p_z}}&0&{ - m{c^2}}
\end{array}} \right]
\end{array}
\end{equation}
\end{small}
with the basis space
\begin{small}$\{ \left| {10} \right\rangle ,\left| {11} \right\rangle \}
\otimes \{ \left| {20} \right\rangle ,\left| {21} \right\rangle \}
\otimes \{ \left| {{n_x}} \right\rangle \}  \otimes \{ \left| {{n_y}} \right\rangle \}
 \otimes \{ \left| {{n_z}} \right\rangle \} $\end{small}
\par The zitterberweng effect is observed according to the formula~\cite{1}
\begin{small}
\begin{equation}
r(t)\!=\!r(0)\!\!+\! c^2 pH_I^{ - 1}t\!+ \!\frac{c\alpha\!\!-\!\!c^2 pH_I^{-1}}{{2iH_I }}(e^{2iH_I t}\!-\!1)
\end{equation}
\end{small}
In 2+1D, \begin{small}$p_z=0$\end{small} in equation (6). Certainly, 1+1 Dirac equation is reduced equation (6),
${H_I} = m{c^2}{{\hat \sigma }_z} + c{p_x}{{\hat \sigma }_x}$
and 1+1D tremor is observed according to reduced formula (7).
\begin{figure}[tbp]
  \includegraphics[width=3in,height=1.7in]{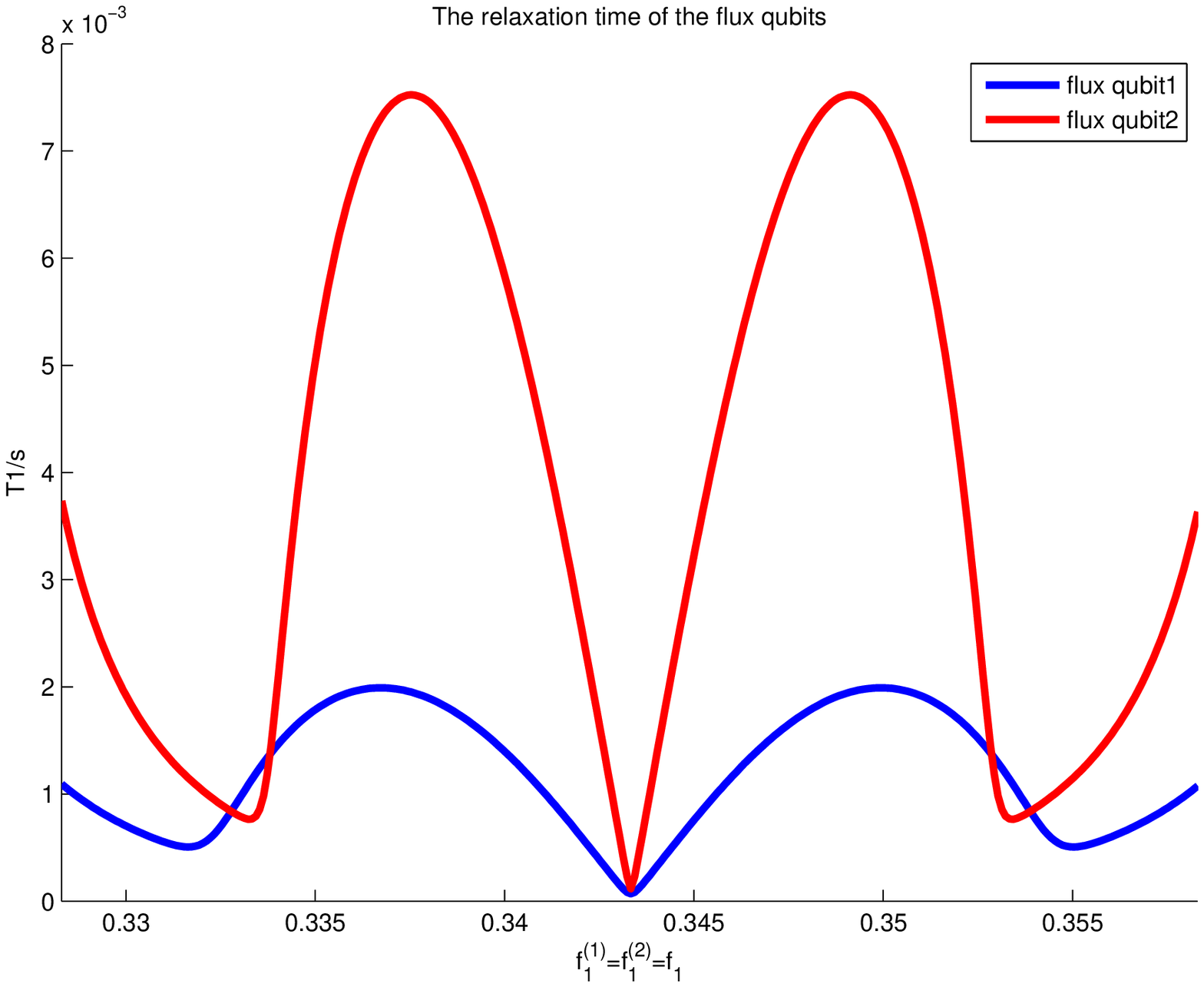}
  \includegraphics[width=3in,height=1.7in]{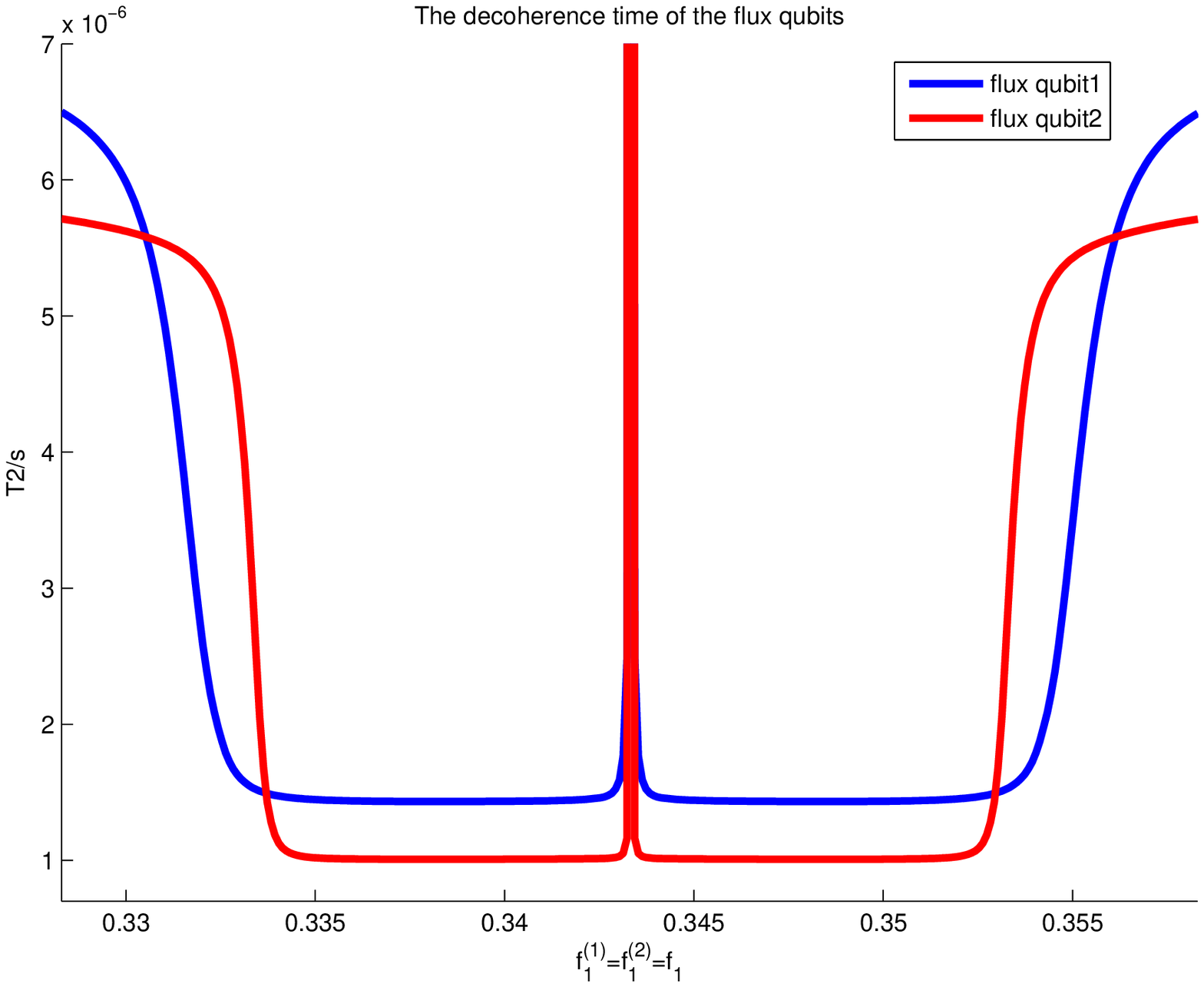}\\
  \includegraphics[width=3in,height=1.7in]{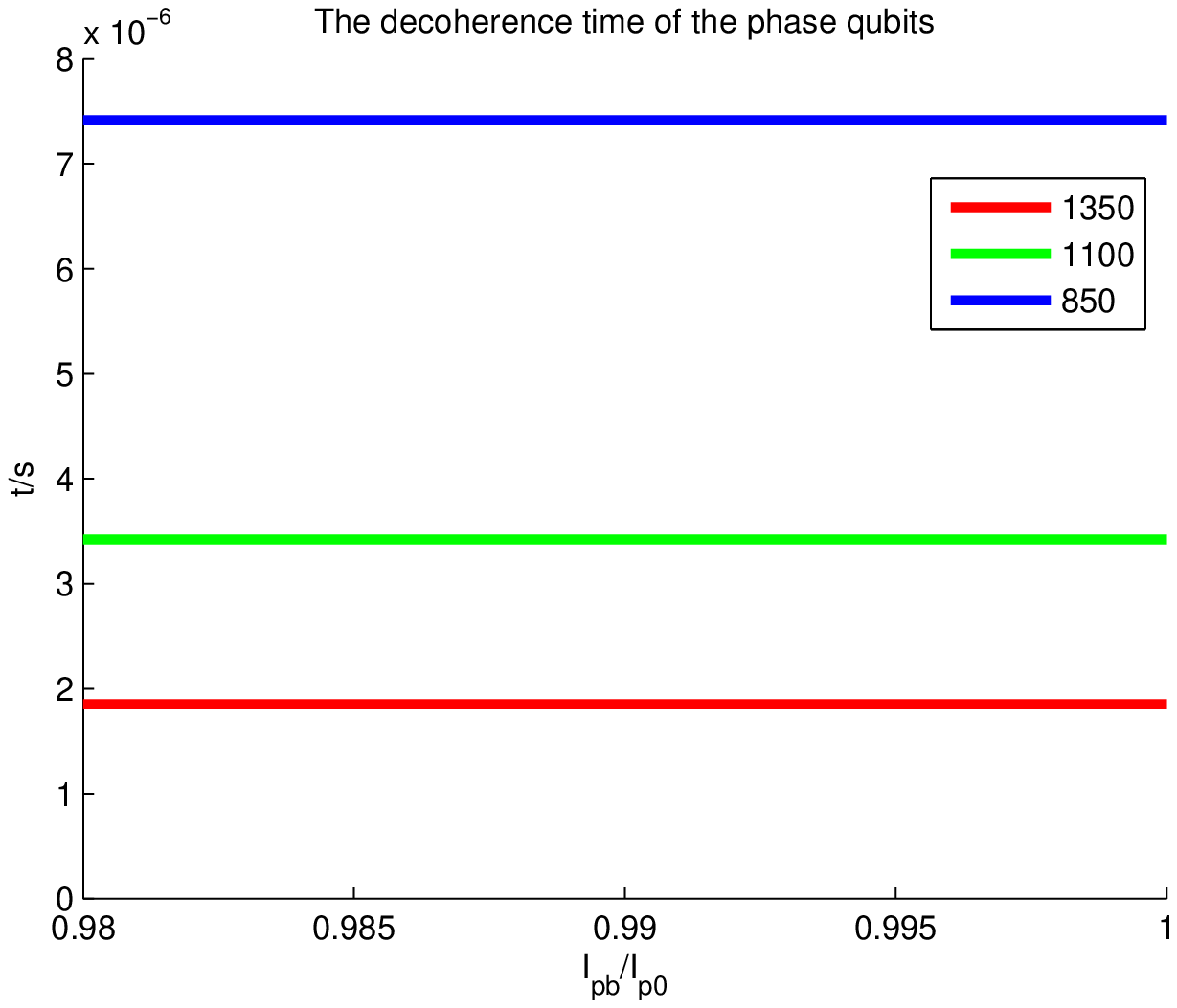}
  \includegraphics[width=3in,height=1.7in]{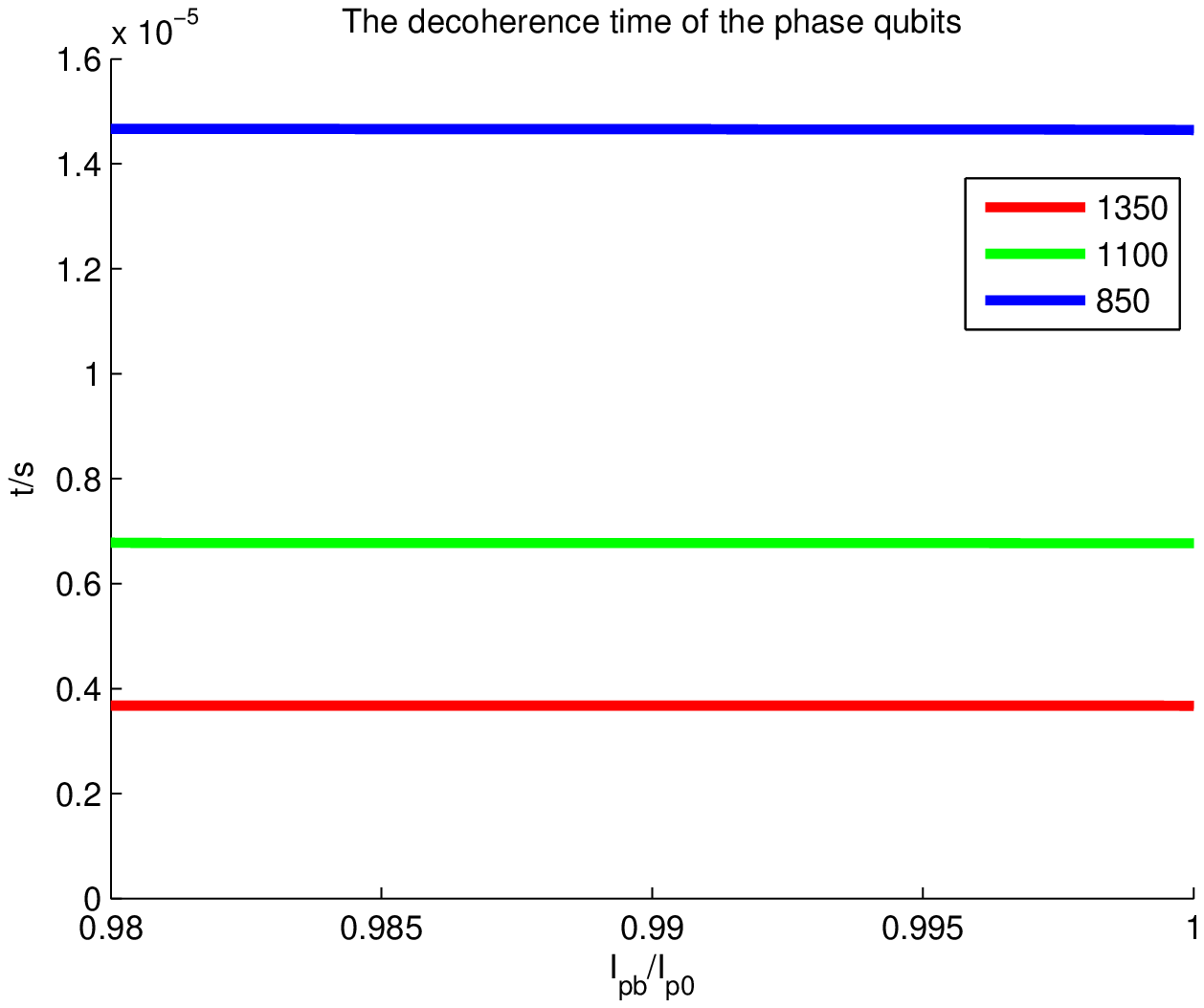}
\caption{the relaxation time and decoherence time of the qubits.
The left graph is relaxation time and the right graph is decoherence time.
The first depict the magnetic flux qubits, voltage bias capacitance $C_g^{(l)}=0$,
magnetic bias inductance $m\sim5 pH$. The second column depict
phase qubits with $E_{Jp}/\hbar=850,\; 1100,\; 1350$GHz,
voltage bias capacitance satisfy $C_{gp}/(C_{Jp}+C_{sh})\sim0$,
magnetic bias inductance $m\sim5 pH$}
\label{}
\end{figure}
\begin{figure}[tbp]
\includegraphics[width=3in,height=1.7in]{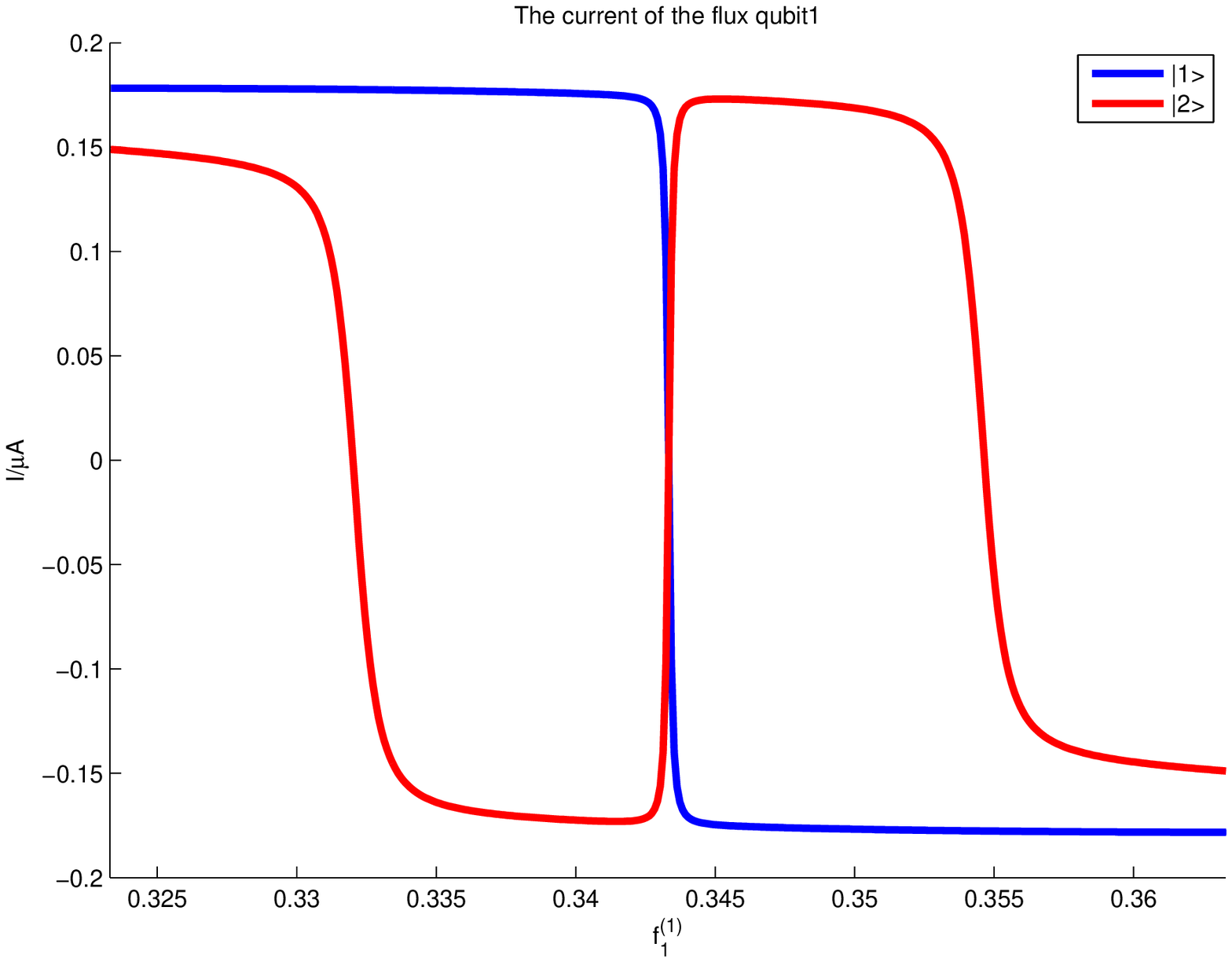}
\includegraphics[width=3in,height=1.7in]{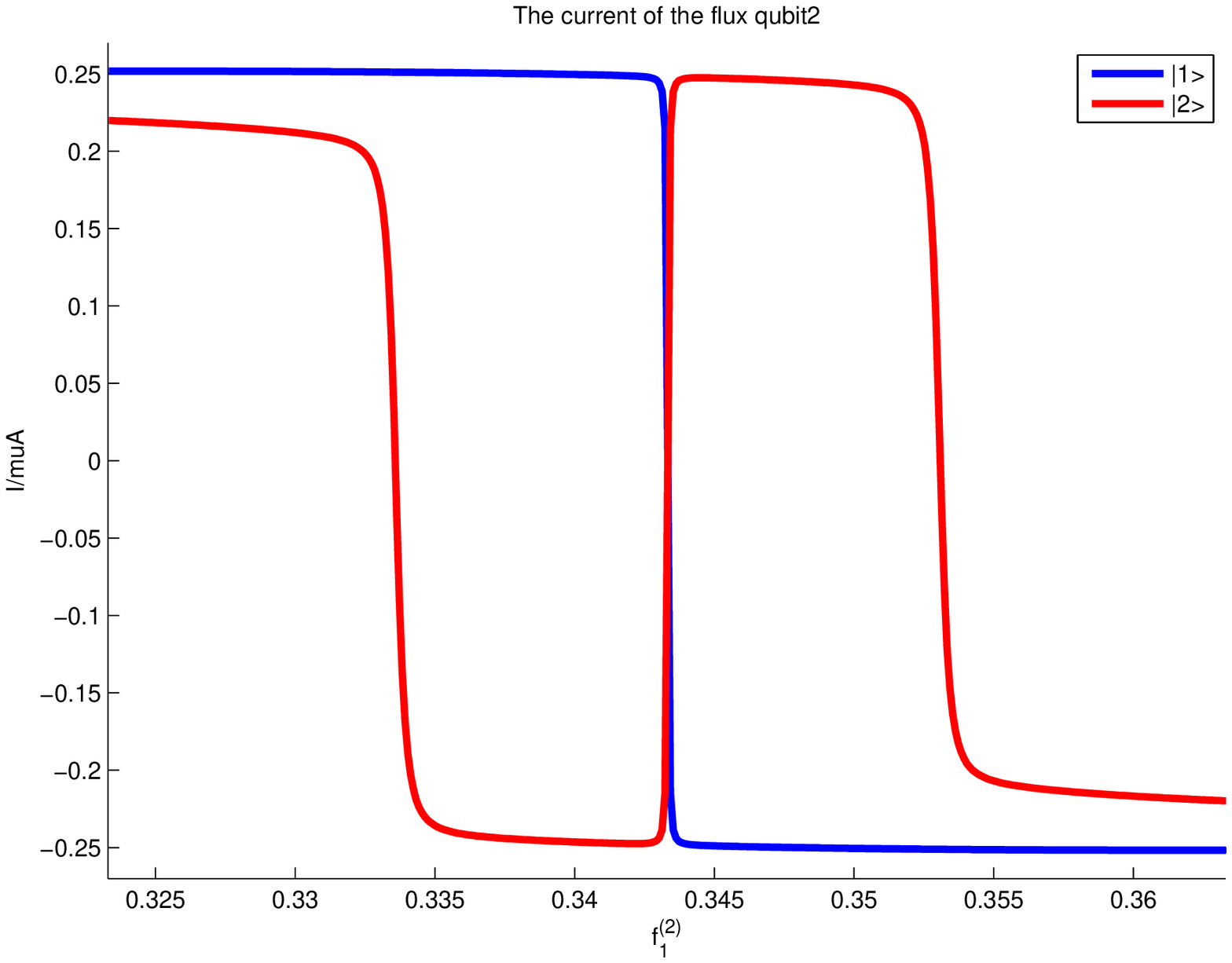}
\includegraphics[width=3in,height=1.7in]{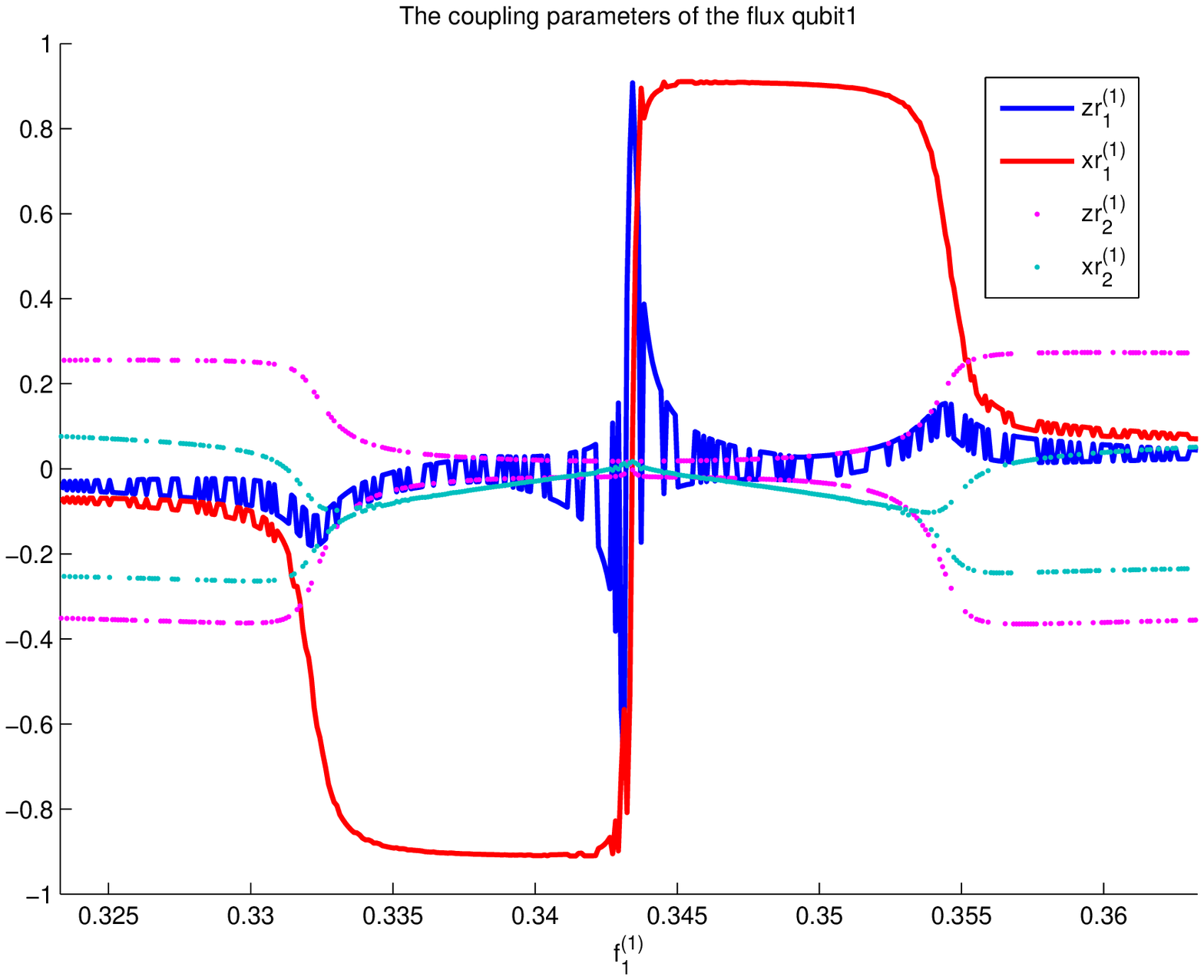}
\includegraphics[width=3in,height=1.7in]{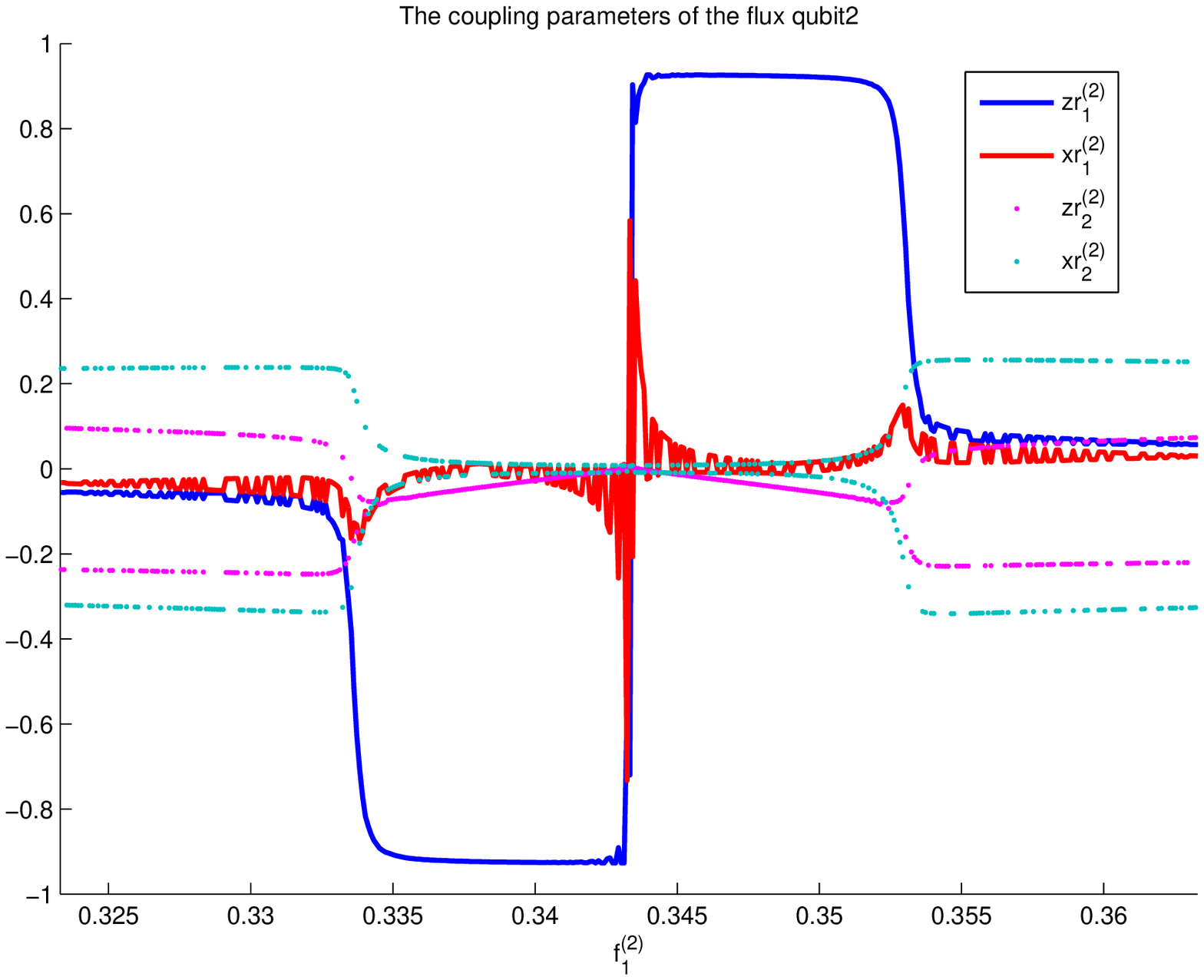}
\caption{The magnetic flux qubits' circling current and
the coupling parameters($zr_i^{(l)}$,$xr_i^{(l)}$,$l=1,2$;$i=1,2$).
Ground state, $\left| {0} \right\rangle$, excited state, $\left| {1} \right\rangle$.
The coupling parameters satisfy
$Z_1^{(l)} = {{\mu}_l}zr_1^{(l)}$,
$X_1^{(l)} = {{\mu}_l}xr_1^{(l)}$,
$Z_2^{(l)} = {{\mu}_l}zr_2^{(l)}$,
$X_2^{(l)} = {{\mu}_l}xr_2^{(l)}$,
${{\mu}_l} = 2{\alpha _l}E_J^{(l)}\cos (\pi f_1^{(l)})$
}
\label{*}
\end{figure}
\par
Quantum states stored at the superconductivity quantum system have been dead accidentally
from thermal noise~\cite{33} or $1/f$ noise and geometric
phase~\cite{35,36,37} prior to the decay naturally due to damping.
Fig.\,3 represents the decoherence time of each qubit in the system.
Fig.\,4 shows some other properties of the magnetic flux qubits, which related to decoherence.
The flux qubits' scheme in Fig 1 is composite of schemes in reference~\cite{20,21,22,23}.
The new flux qubit has a suitable shunt capacitor, a thinner barrier determined
by the SQUID set at proper operating point, a lager ring inductor. These new features
effectively reduces the dephasing of the charge noise, the magnetic noise,
and effectively broadens the stable operating region of flux qubit~\cite{20,21,22,23}.
Considering the AC component in the circling current, the coupling parameters of the flux qubit are fluctuating.
When \begin{small}$f_1^{(l)}\sim0.3310$ or $0.3555$\end{small}, the circling current is smaller,
\begin{small}$Z_1^{(l)}$, $X_1^{(l)}$\end{small} are smaller,
\begin{small}$Z_2^{(l)}$, $X_2^{(l)}$\end{small} are larger and stable.
So, the operating point shift of flux qubit is smaller due to transition of adjacent flux qubit.
For $\sim MHz$ noise, the adiabatic condition is well met~\cite{35,36,37}, the decoherence effects of the driving
pulses are completely ignore. The coupling strength mapping to the kinetic energy is lager.
For the phase qubits, a single crystal silicon shunt capacitor have released the relaxation time~\cite{38}.
When operating point at \begin{small}$I_{pb}/I_{p0}=0.99$\end{small}, \begin{small}$\lambda_p$ is larger\end{small}.
\par In the current low temperature conditions, about $6mK$,
there is only a small influence of thermal fluctuation and photon noise~\cite{33,41}.
The thermal annealing has also enhanced dephasing time~\cite{49}.
Nondestructive measurement reduces the damage from readout~\cite{51}.
In our system, the ideal expected results can be obtained if the decoherence time
can reach a microsecond. Resonance does not contribute to decoherence~\cite{52},
then the shortest decoherence time among all the qubits under the strict conditions
is closer to the lower bound of the decoherence time of the system.
\par In conclusion, we have designed experimentally feasible quantum simulators to
simulate 3+1, 2+1, 1+1 Dirac equation.
\par We are grateful to H.Q. Lin, J.Q. You, Jiuqing Liang, Changjun Liao
for fruitful discussions.

\appendix
Appendix: Details of the iterative calculations\\
Part A: Quantization of the flux qubit\\\\
Part B: Quantization of the phase qubit\\\\
Part C: calculation of decoherence time\\\\
Part D: Details of the iteration calculation\\\\
From the view of quantum theory, we calculate the coupling
between three qubits driven resonantly by the driving pulses and prove the
inductance of the josephson junction is $\Phi _0^2/(4{\pi ^2}{E_J})$, which is consistent with
the classic conclusion. To be simplified, $\hat a^ -$
represents annihilation operator, Einstein summation symbol is applied.
$"\pm"$ is the same in the following cases: $\hat\sigma_\pm ^{(2)}$ and $\pm 2{\varepsilon _2}$;
$\hat a_x^ \pm$ and $\pm {\omega _x}$; $\pm \omega _i^{(1)}$ and $\pm \phi _i^{(1)}$;
$\pm \omega _i^{(2)}$ and $\pm \phi _i^{(2)}$;
The left $\hat a_P^\pm$ in the $\hat a_P^\pm\hat a_P^\pm$ and the left $\pm {\omega _P}$
in the $\pm{\omega _P}\pm{\omega _P}$ in the exponential expression
and the leftmost $\pm {\omega _P}$ in the denominator;
The right $\hat a_P^\pm$ in the $\hat a_P^\pm\hat a_P^\pm$ and the right $\pm {\omega _P}$
in the $\pm{\omega _P}\pm{\omega _P}$ in the exponential expression
and the other $\pm {\omega _P}$ in the denominator.
\\
\par Part A: Quantization of the flux qubit~\cite{23,25,26,27,29,31}
\\
\par
When ring inductance ${L_r} \to 0$, the Hamiltonian of the system is shown as\cite{31}
\begin{small}
\[\begin{array}{l}
H = \sum\limits_{i = 1}^4 {[{{({\Phi _0}/2\pi )}^2}{C_i}\dot \varphi _i^2/2 + {E_{Ji}}(1 - \cos {\varphi _i})]}+L_r I^2/2 \\
 = {\raise0.5ex\hbox{$\scriptstyle 1$}
\kern-0.1em/\kern-0.15em
\lower0.25ex\hbox{$\scriptstyle 2$}}C\left[ {\begin{array}{*{20}{c}}
{{Q_1}}&{{Q_2}}
\end{array}} \right]\left[ {\begin{array}{*{20}{c}}
{1 + \beta }&\beta \\
\beta &{1 + \beta }
\end{array}} \right]\left[ {\begin{array}{*{20}{c}}
{{Q_1}}\\
{{Q_2}}
\end{array}} \right]\!\!-\!\!{E_J}\cos {\varphi _1}\!\!-\!\!{E_J}\cos {\varphi _2}\!\!-\!\!
2\alpha {E_J}\cos (\pi {f_2})\cos ({\varphi _1}\!\!+\!\!{\varphi _2} + 2\pi {f_1} + \pi {f_2})\\
 = \left[ {\begin{array}{*{20}{c}}
{{Q_1}}&{{Q_2}}
\end{array}} \right]\left[ {\begin{array}{*{20}{c}}
{\frac{{1 + \beta }}{{2(1 + 2\beta )C}}}&{\frac{{ - \beta }}{{2(1 + 2\beta )C}}}\\
{\frac{{ - \beta }}{{2(1 + 2\beta )C}}}&{\frac{{1 + \beta }}{{2(1 + 2\beta )C}}}
\end{array}} \right]\left[\!{\begin{array}{*{20}{c}}
{{Q_1}}\\
{{Q_2}}
\end{array}}\!\right]\!\!-\!\!{E_J}\cos {\varphi _1}\!\!-\!\!{E_J}\cos {\varphi _2}\!\!-\!\!
2\alpha {E_J}\cos (\pi {f_2})\cos ({\varphi _1}\!\!+ {\varphi _2}\!\!+\!\!2\pi {f_1} + \pi {f_2})\\
 = \frac{{(1 + \beta )Q_1^2}}{{2(1 + 2\beta )C}} + \frac{{(1 + \beta )Q_2^2}}{{2(1 + 2\beta )C}} - \frac{{\beta {Q_1}{Q_2}}}{{(1 + 2\beta )C}}
\!\!-\!\!{E_J}\cos {\varphi _1}\!\!-\!\!{E_J}\cos {\varphi _2}\!\!-\!\!2\alpha {E_J}\cos (\pi {f_2})\cos ({\varphi _1}\!\!+\!\!
{\varphi _2}\!\!+\!\!2\pi {f_1}\!+\!\pi {f_2})
\end{array}\]
\end{small}
the first derivative of the phase $\varphi _1$ is shown as
\begin{small}
\[{\dot \varphi _1}\!=\!\frac{i}{\hbar }[H,{\varphi _1}]\!=\!\frac{i}{\hbar }\frac{{2e}}{i}\frac{{(1\!+\!\beta )}}{{(1\!+\!2\beta )C}}\frac{{2e}}{i}\frac{\partial }{{\partial {\varphi _1}}}\!+\!\frac{i}{\hbar }\frac{{2e}}{i}\frac{{ - \beta }}{{(1 + 2\beta )C}}{Q_2}\!=\!\frac{{2\pi }}{{{\Phi _0}}}\frac{{(1 + \beta )}}{{(1 + 2\beta )C}}\frac{\partial }{{\partial {\varphi _1}}}\!+\!\frac{{2\pi }}{{{\Phi _0}}}\frac{{ - \beta }}{{(1 + 2\beta )C}}{Q_2}\]
\end{small}
the second derivative of the phase $\varphi _1$ is shown as
\begin{small}
\[{{\ddot \varphi }_1} = \frac{i}{\hbar }[H,{{\dot \varphi }_1}] =  - {\left( {\frac{{2\pi }}{{{\Phi _0}}}} \right)^2}\frac{1}{C}{E_J}\left[ {\frac{{1 + \beta }}{{1 + 2\beta }}\sin {\varphi _1} + \frac{{ - \beta }}{{1 + 2\beta }}\sin {\varphi _2} + \frac{{2\alpha \cos (\pi {f_2})}}{{1 + 2\beta }}sin({\varphi _1} + {\varphi _2} + 2\pi f)} \right]\]
\end{small}
the current of junction1 in the flux qubit
\begin{small}
\[{I_0} = {I_{c1}}\sin {\varphi _1} + {C_1}\frac{{{\Phi _0}}}{{2\pi }}{\ddot \varphi _1} = \frac{{2\pi }}{{{\Phi _0}}}\frac{\beta }{{1 + 2\beta }}{E_J}\left[ {\sin {\varphi _1} + \sin {\varphi _2} - 2(\alpha /\beta )\cos (\pi {f_2})sin({\varphi _1} + {\varphi _2} + 2\pi f)} \right]\]
\end{small}
the Hamiltonian of the flux qubit with ring inductance $L_r$ by driving pulses represents as
\begin{small}
\[\begin{array}{l}
H_q^{(l)} = 2E_{Ca}^{(l)}N_{al}^2\!+\!2E_{Cs}^{(l)}N_{sl}^2\!+\!2E_J^{(l)}\!+\!2{\alpha _l}E_J^{(l)}\\
\;\;\;\;\;\;\;\;-2E_J^{(l)}\cos \varphi _a^{(l)}\cos \varphi _s^{(l)}-\!2{\alpha _1}E_J^{(l)}
\cos (\pi f_2^{(l)})\cos (2\varphi _s^{(l)}\!+\!2\pi f_3^{(l)})\\
\;\;\;\;\;\;\;\;+2{\alpha _l}E_J^{(l)}\cos (\pi f_2^{(l)})sin(2\varphi _s^{(l)}\!+\!2\pi f_3^{(l)})[\varphi _r^{\scriptscriptstyle{(l)}}\!+\!\sum\limits_{i\!=\!1}^3\!{\Phi _i^{(l)}}]\\
\;\;\;\;\;\;\;\;+{\alpha _l}E_J^{(l)}\cos (\pi f_2^{(l)})\cos (2\varphi _s^{(l)}\!+\!2\pi f_3^{(l)}){[\varphi _r^{\scriptscriptstyle{(l)}}\!\!+\!\!\sum\limits_{i\!=\!1}^3\!{\Phi _i^{(l)}}]^2}\\
\;\;\;\;\;\;\;\;+[2{\beta _1}/(1\!+\!4{\beta _l})]\hbar N_s^{(l)}\sum\limits_{i\!=\!1}^3 {\dot \Phi _i^{(l)}}
\!+\!\frac{1}{2}L{[{I^{(l)}}]^2}
\end{array}\]
\end{small}
\par When ring inductance ${L_r} \ne 0$, the Hamiltonian of the system is shown as
\begin{small}
\[H = \sum\limits_{i = 1}^4 {[{{({\Phi _0}/2\pi )}^2}{C_i}\dot \varphi _i^2/2 + {E_{Ji}}(1 - \cos {\varphi _i})]}
 + {\textstyle{1 \over 2}}L{(2\pi {E_{J1}}/{\Phi _0})^2}{\sin ^2}{\varphi _1}\]
\end{small}
The circling current is approximated as
\begin{small}
\[{I_0} = \frac{{2\pi }}{{{\Phi _0}}}\frac{\beta }{{1 + 2\beta }}{E_J}\left[ {\sin {\varphi _1} + \sin {\varphi _2} - 2(\alpha /\beta )\cos (\pi {f_2})sin({\varphi _1} + {\varphi _2} + 2\pi f) + l\sin {\varphi _1}\cos {\varphi _1}} \right]\]
\end{small}
Here, $l=L_r/L_J$, $L_r$ ring inductance, $L_J$ junction inductance.
\\Numerical solution shows that there is little difference in the two cases with $l=0.17;\;l=0.23$.
\\The Hamiltonian of the bare flux qubit biased static magnetic fluxes represents as\cite{31}
\begin{small}
\[\begin{array}{l}
H_{q0}^{(l)}\!=\!2E_{Ca}^{(l)}N_{al}^2\!+\!2E_{Cs}^{(l)}N_{sl}^2\!-\!2E_J^{(l)}\!\cos\!\varphi _s^{(l)}\!\cos\!\varphi _a^{(l)}
 \!\!-\!\!2{\alpha _l}E_J^{(l)}\!\cos\!(\pi f_2^{(l)})\!\cos\!(2\varphi _s^{(l)}\!+\!2\pi f_3^{(l)})\!-\!L_r^{(l)}{(I_0^{(l)})^2}/2
\end{array}\]
\end{small}
corresponding circling current represents as
\begin{small}
\[\begin{array}{l}
I_0^{(l)} = [{\beta _l}/(1 + 2{\beta _l})](2\pi /{\Phi _0})E_J^{(l)}[2\sin \varphi _s^{(l)}\cos \varphi _a^{(l)}
- (2{\alpha _l}/{\beta _l})\cos (\pi f_2^{(l)})\sin (2\varphi _s^{(l)} + 2\pi f_3^{(l)})]
\end{array}\]
\end{small}
\[H_{q0}^{(l)} = z_0^{(l)}\hat \sigma _z^{(l)} + x_0^{(l)}\hat \sigma _x^{(l)}\]
the Hamiltonian of $H_q^{(l)}$ is transformed into
\[\begin{array}{l}
H_D^{(l)} = D_l^T({\theta _l}/2)H{D_l}({\theta _l}/2)\;,\;\;{\theta _l} = \arccos (z_0^{(l)}{[{(z_0^{(l)})^2} + {(x_0^{(l)})^2}]^{1/2}}),\;\\\\
{D_1}({\theta _1}) = \left[ {\begin{array}{*{20}{c}}
{\cos (\pi /4 + {\theta _1}/2)}&{sin(\pi /4 + {\theta _1}/2)}\\
{sin(\pi /4 + {\theta _1}/2)}&{ - \cos (\pi /4 + {\theta _1}/2)}
\end{array}} \right]\;,\;\;\;{D_2}({\theta _2}) = \left[ {\begin{array}{*{20}{c}}
{\cos ({\theta _2}/2)}&{\sin ({\theta _2}/2)}\\
{\sin ({\theta _2}/2)}&{\cos ({\theta _2}/2)}
\end{array}} \right]
\end{array}\]
$H_{q0}^{(1)}$, $H_{q0}^{(2)}$ are transformed into
\[H_{D0}^{(1)} = X_0^{(1)}\hat \sigma _x^{(1)},\;\;H_{D0}^{(2)} = Z_0^{(2)}\hat \sigma _z^{(2)}\]
\\\\
Part B: the quantization of the phase qubit:
\\\\
Define the phase
\[{\varphi _{p1}} = {\varphi _p} - {\varphi _{p0}},\;\]
\[{\gamma _{P1}} = {\gamma _P} - {\gamma _{P0}}\;,\]
bias current
\[\sin {\varphi _{p0}} = {I_{pb}}/{I_{p0}}\;,\]
the phase relationship:
\[{\varphi _{p0}} - {\gamma _{P0}} + 2\pi {\Phi _p}/{\Phi _0} = 0\;,\;\;\]
the effective coupling energy of the inductance $L_{rp}$
\[{E_{rp}} = {({\Phi _0}/2\pi )^2}/{L_{rp}},\]
The Hamiltonian of the phase qubit $p$
\[\begin{array}{l}
{H_p} = 4{E_{cp}}N_p^2 - {E_{Jp}}(\cos {\varphi _p} + {I_{pb}}{\varphi _p}/{I_{p0}}) + {\textstyle{1 \over 2}}{E_{rp}}{({\varphi _p} - {\varphi _{p0}})^2}\\
\\
\;\;\;\;\; = 4{E_{cp}}N_p^2 - {E_{Jp}}[\cos {\varphi _{p1}}\cos {\varphi _{p0}} - \sin {\varphi _{p1}}\sin {\varphi _{p0}} + {\varphi _{p1}}sin{\varphi _{p0}} + {\varphi _{p0}}sin{\varphi _{p0}}] + {\textstyle{1 \over 2}}{E_{rp}}{({\varphi _p} - {\varphi _{p0}})^2}\\
\\
\;\;\;\;\; = 4{E_{cp}}N_p^2 - {E_{Jp}}[\cos {\varphi _{p1}}\cos {\varphi _{p0}} - {\varphi _{p1}}\sin {\varphi _{p0}} + {\varphi _{p1}}sin{\varphi _{p0}} + {\varphi _{p0}}sin{\varphi _{p0}}] + {\textstyle{1 \over 2}}{E_{rp}}{({\varphi _p} - {\varphi _{p0}})^2}\\
\\
\;\;\;\;\; = 4{E_{cp}}N_p^2 - {E_{Jp}}[\cos {\varphi _{p1}}\cos {\varphi _{p0}} + {\varphi _{p0}}sin{\varphi _{p0}}] + {\textstyle{1 \over 2}}{E_{rp}}{({\varphi _p} - {\varphi _{p0}})^2}\\
\\
\;\;\;\;\; = 4{E_{cp}}N_{p1}^2 + {\textstyle{1 \over 2}}{E_{Jp}}\cos {\varphi _{p0}}\varphi _{p1}^2 + {\textstyle{1 \over 2}}{E_{rp}}{({\varphi _p} - {\varphi _{p0}})^2}\\
\\
\;\;\;\;\; = 4{E_{cp}}N_{p1}^2 + {\textstyle{1 \over 2}}({E_{Jp}}\cos {\varphi _{p0}} + {E_{rp}})\varphi _{p1}^2\;
\end{array}\]
In the population represention:
\[{\varphi _{p1}} = {\lambda _p}(\hat a_p^ +  + {{\hat a}_p}),{N_p} = i(\hat a_p^ +  - {{\hat a}_p})/2{\lambda _p}\;,\]
The boson operator
\[\hat a_p^ +  = {\varphi _{p1}}/(2{\lambda _p}) - i{\lambda _p}{N_p}\;,\;{{\hat a}_p} = {\varphi _{p1}}/(2{\lambda _p}) + i{\lambda _p}{N_p}\]
\[{\lambda _p} = {[2{E_{Cp}}/({E_{Jp}}\cos {\varphi _{p0}} + {E_{rp}})]^{1/4}},\;\hbar {\omega _p} = {[8{E_{Cp}}({E_{Jp}}\cos {\varphi _{p0}} + {E_{rp}})]^{1/2}}\]
the Hamiltonian is secondary quantized to
\[{H_p} = \hbar {\omega _p}(\hat a_p^ + {{\hat a}_p} + 1/2)\]
Total Hamiltonian is described as
\begin{small}
\[\begin{array}{l}
{H_I} = \hbar X_0^{(1)}\hat \sigma _x^{(1)}\!\!+\!\!\hbar Z_0^{(2)}\hat \sigma _z^{(2)}\!\!+\!\!\sum\limits_{P = C} {\hbar {\omega _P}\hat a_P^ + {{\hat a}_P}}  \!\!+\!\!\sum\limits_{p = c} {\hbar {\omega _p}\hat a_p^ + {{\hat a}_p}}\!\!+\!\!\hbar {\omega _O}\hat a_O^ + {{\hat a}_O}\\
 + (\hbar Z_1^{(l)}\hat \sigma _z^{(l)} + \hbar X_1^{(l)}\hat \sigma _x^{(l)})[({( - 1)^l}{\lambda _O}\hat a_O^ +\!\!+\!\!\sum\limits_{{P_l} = {C_l}} {{\lambda _{{P_l}}}\hat a_{{P_l}}^ + }\!\!+\!\!\sum\limits_{r = {R_l}} {{\textstyle{1 \over 2}}n_r^{(l)}{e^{\omega _r^{(l)}it + i\phi _r^{(l)}}}) + H.C} ]\\
 + (\hbar Z_2^{(l)}\hat \sigma _z^{(l)} + \hbar X_2^{(l)}\hat \sigma _x^{(l)})[{({( - 1)^l}{\lambda _O}\hat a_O^ +\!\!+\!\!\sum\limits_{{P_l} = {C_l}} {{\lambda _{{P_l}}}\hat a_{{P_l}}^ + }  +\!\!\sum\limits_{r = {R_l}} {{\textstyle{1 \over 2}}n_r^{(l)}{e^{\omega _r^{(l)}it + i\phi _r^{(l)}}}) + H.C} ]^2}\\
 + (\hbar Z_3^{(l)}\hat \sigma _z^{(l)} + \hbar X_3^{(l)}\hat \sigma _x^{(l)})\sum\limits_{r = {R_l}}\!\!{{\textstyle{1 \over 2}}n_r^{(l)}\omega _r^{(l)}({e^{\omega _r^{(l)}it + i(\phi _r^{(l)} - \pi /2)}} + H.C)} \\
 + {E_{Lp}}{[\sum\limits_{pP} {({\lambda _p}\hat a_p^ +  - {\lambda _P}\hat a_P^ + )}  + H.C]^2}
\end{array}\]
\end{small}
\begin{small}
\[\begin{array}{l}
3 + 1D:\;C = X,Y,Z;c = x,y,z;{R_l} = 3;l = 1:{C_l} = X,Z;p = x,z;\;l = 2:{C_l} = Y;\;p = y;\\
\;\;\;\;\;\;\;\;\;\;\;\;\;\;pP:p = x\& P = X;p = y\& P = Y;p = z\& P = Z;\\
2 + 1D:C = X,Y,T;c = x,y,z;l = 1:{C_l} = X;p = x;\;{R_l} = 3;l = 2:{C_l} = Y;p = y;{R_l} = 2\\
\;\;\;\;\;\;\;\;\;\;\;\;\;\;pP:p = x\& P = X;p = y\& P = Y;\\
1 + 1D:C = X;c = x;\;{R_l} = 2;{\lambda _O} = 0;\;l = 1:{C_l} = X;pP:p = x\& P = X;
\end{array}\]
\end{small}
the factors in the Hamiltonian are shown as
\[\begin{array}{l}
Z_i^{(l)} = {m_l}zr_i^{(l)}\hat \sigma _z^{(l)},\;\;\;X_i^{(l)} = {m_l}xr_i^{(l)}\hat \sigma _x^{(l)}\\
when\;\;i = 0,1,2,\;\;{m_l} = {\mu _l};\;\;i = 3,\;\;{m_l} = {\nu _l};\;l = 1,\;2\\
{\mu _l} = 2{\alpha _l}E_J^{(l)}\cos (\pi f_1^{(l)}),\;\;\;\;{\nu _l} = 2{\beta _l}/(1 + 4{\beta _l})\\
{\theta _l} = \arccos (z_0^{(l)}{[{(z_0^{(l)})^2} + {(x_0^{(l)})^2}]^{ - 1/2}})\;\\
zr_i^{(1)} = x_i^{(1)}\cos {\theta _1} - z_i^{(1)}\sin {\theta _1},\;\;xr_i^{(1)} = x_i^{(1)}\sin {\theta _1} + z_i^{(1)}\cos {\theta _1}\\
zr_i^{(2)} = z_i^{(2)}\cos {\theta _2} + x_i^{(2)}\sin {\theta _2},\;xr_i^{(2)} = z_i^{(2)}\sin {\theta _2} - x_i^{(2)}\cos {\theta _2}\\
z_0^{(l)} = (\langle {e_l}|H_{q0}^{(l)}|{e_l}\rangle  - \langle {g_l}|H_{q0}^{(l)}|{g_l}\rangle )/2\;,\;x_0^{(l)} = \langle {e_l}|H_{q0}^{(l)}|{g_l}\rangle \;,\\
z_1^{(l)} = (\langle {e_l}|si{n_l}|{e_l}\rangle  - \langle {g_l}|si{n_l}|{g_l}\rangle )/2\;,\;x_1^{(l)} = \langle {e_l}|si{n_l}|{g_l}\rangle \\
z_2^{(l)} = (\langle {e_l}|co{s_l}|{e_l}\rangle  - \langle {g_l}|co{s_l}|{g_l}\rangle )/2,\;x_2^{(l)} = \langle {e_l}|co{s_l}|{g_l}\rangle \;\\
{\sin _l} = sin(2\varphi _a^{(l)} + 2\pi f_3^{(l)})\;,\;{\cos _l} = \cos (2\varphi _a^{(l)} + 2\pi f_3^{(l)})\\
z_3^{(l)} = (\langle {e_l}|{P_{al}}|{e_l}\rangle  - \langle {g_l}|{P_{al}}|{g_l}\rangle )/2\;,\;\;x_3^{(l)} = \langle {e_l}|{P_{al}}|{g_l}\rangle
\end{array}\]
\\
\par Part C: calculation of decoherence time
\\
\par The decoherence time is calculated as
\begin{small}
\[\begin{array}{l}
{\Gamma _1} = {\sum\nolimits_i {|{B_i}|} ^2}{S_i}({\omega _{10}}) = \frac{{{{\left\langle {{Q_i}} \right\rangle }^2}}}{{{\hbar ^2}}}{\left( {\frac{{{C_g}}}{C}} \right)^2}\hbar (2\pi {\omega _{10}})Re[Z({\omega _{10}})] + \frac{{{{\left\langle I \right\rangle }^2}}}{{{\hbar ^2}}}{M^2}\hbar (2\pi {\omega _{10}})Re[Y({\omega _{10}})]\\
{\Gamma _\varphi } = \sum\nolimits_i {|{A_i}|} {[{\alpha _i}ln({\omega _t}/{\omega _c})]^{0.5}}\\
{\Gamma _2} = {\Gamma _1}/2 + {\Gamma _\varphi }\\
{T_1} = 1/{\Gamma _1}\\
{T_2} = 1/{\Gamma _2}
\end{array}\]
Here, $Q_i$ is charge in junction capacitor,$I$ is the circling current, ${{C_g}}$ is gate capacitance, $C$ is junction capacitance, $M$ is bias inductance.
In our scheme, no voltage source is applied, so ${{C_g}}=0$.
\end{small}
\par For the frequency of driving field \begin{small}$\omega _i^{(l)}>5 GHz$\end{small},
the geometric dephasing is caused low frequency noise in the driving field
rather than the driving field itself~\cite{35,36,37}.
\begin{figure}[tbp]
\includegraphics[width=2in]{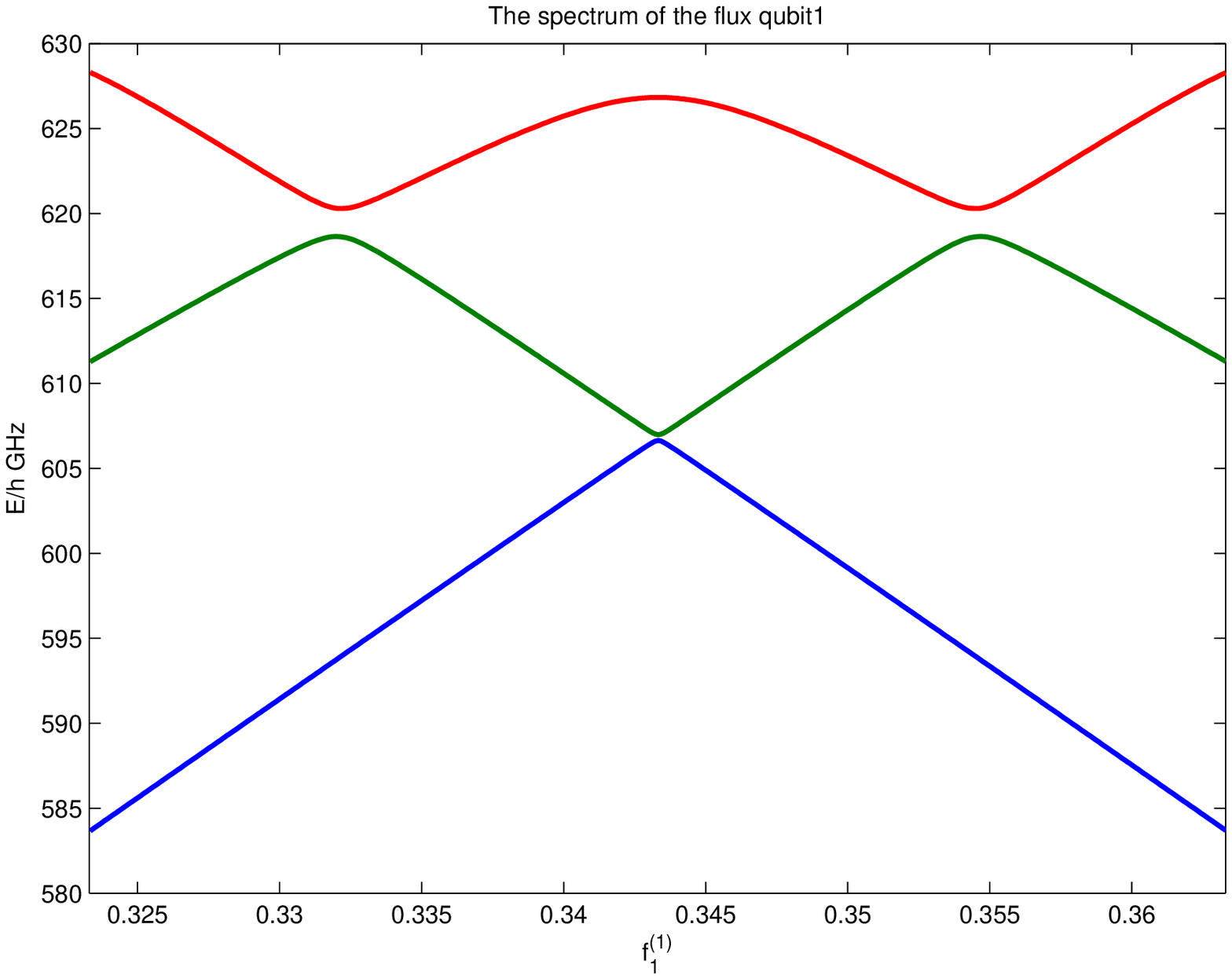}
\includegraphics[width=2in]{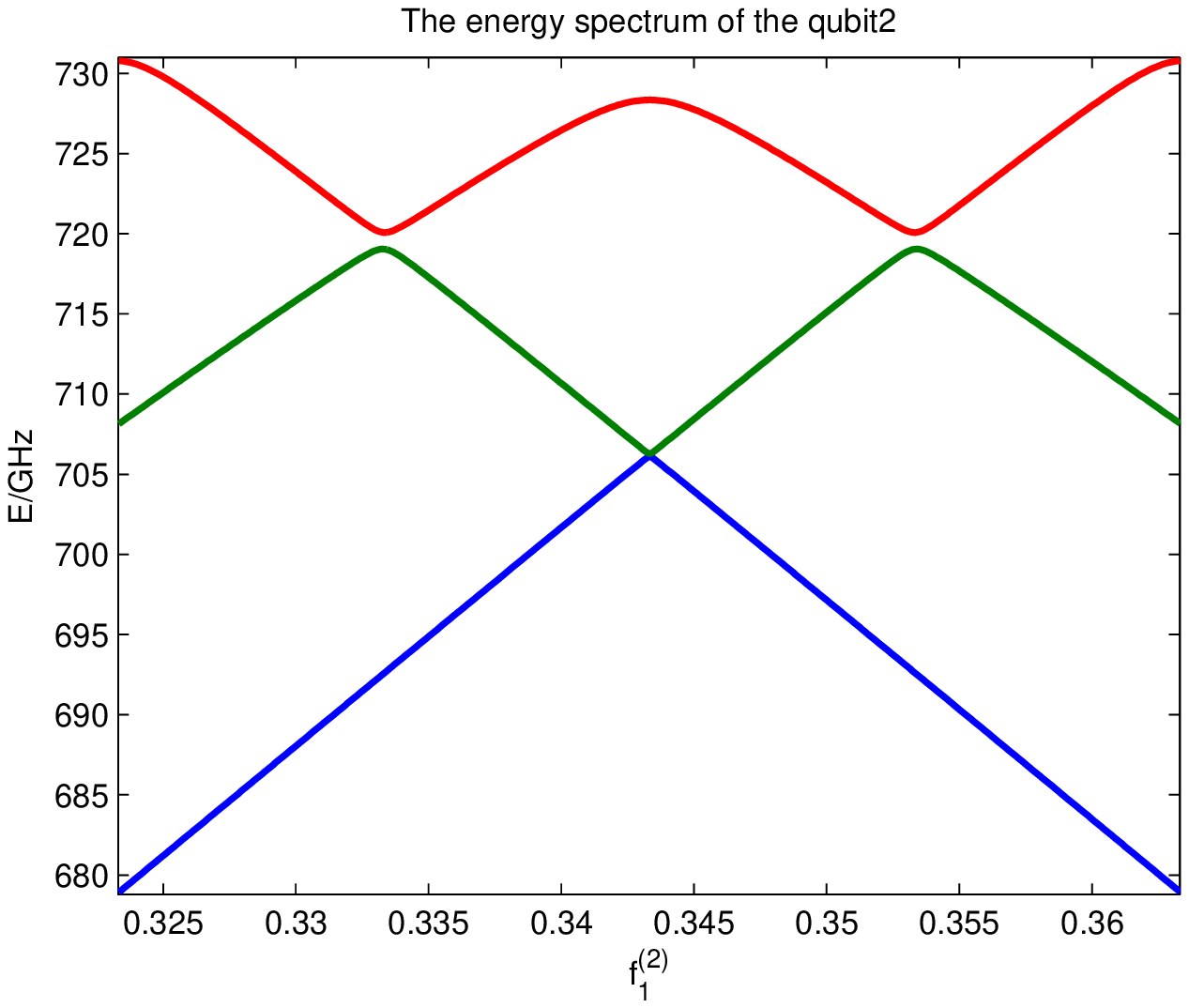}
\caption{The magnetic flux qubits' energy spectrum}
\label{*}
\end{figure}
The detuning \begin{small}${\Delta _i^{(1)}} = 2X_0^{(1)} - \omega _i^{(1)}>5GHz$\end{small},
\begin{small}${\Delta _i^{(2)}} = 2Z_0^{(1)} - \omega _i^{(2)}>5GHz$\end{small}, the coupling strength between
the flux qubit and driving field
\begin{small}$\hbar \Omega_i^{(1)} =\hbar {[{(Z_1^{(1)})^2} + {(Z_3^{(1)}{\omega _3})^2}]^{1/2}}{n_i^{(1)}}$\end{small},
\begin{small}$\hbar \Omega_i^{(2)} =\hbar {[{(X_1^{(2)})^2} + {(X_3^{(2)}{\omega _3})^2}]^{1/2}}{n_i^{(2)}}/2$\end{small}.
According to Fig4, \begin{small}$\Omega _i^{(l)}/{[{(\Delta _i^{(l)})^2} + {(\Omega _i^{(l)})^2}]^{1/2}} < 0.1$\end{small}.
For $\sim MHz$ noise, the adiabatic condition is well met~\cite{37},
which means the geometric phase is very small but exists.
Inspired by spin echo~\cite{37}, the dephasing effects of driving
fields with opposite phase are counteracted each other.
The dephasing effects are also counteracted by low frequency fields induced~\cite{35}.
\\
\\
\\
\\
Part D: Details of the iteration calculation
\\
\par In 3+1D, the first order
approximation is resonant coupling term between
the flux qubit1 and weak time-dependent magnetic fluxes,
corresponding to the rest mass of the particles. It is shown as
\begin{small}
${H_{I1}} = \hbar Z_1^{(1)}n_3^{(1)}\sigma _z^{(1)} +\hbar {\omega _3}Z_3^{(1)}n_3^{(1)}\sigma _y^{(1)}$
\end{small}.
The second order approximation has been considered because it is related to self-coupling of qubit.
The third order approximation is the resonant coupling term between two flux qubits, one phase qubit and
weak time-dependent magnetic fluxes, corresponding to the momentum-spin coupling of the particles.
It can be obtained by iteration of
\begin{small}
${H_{I3}} = \sum {{I_j}\int_0^{{t_1}} {{I_k}d{t_2}\int_0^{{t_2}} {{I_m}d{t_3}} } }$
\end{small},
where \begin{small}$j,k,m=1,2,3$\end{small} and \begin{small}$j \ne k \ne m$\end{small}~\cite{23,25,26,27,29,31}.
All calculations satisfy the conditions of the frequencies:
\begin{small}
\[\begin{array}{l}
 + 2X_0^{(1)} - \omega _3^{(1)} = 0,\;\;\phi _3^{(1)} = 0,\;n_3^{(1)} = {n_m}\\
 + {\omega _z} - \omega _3^{(2)} = 0,\;\phi _3^{(2)} =  - \pi /2,\;n_3^{(2)} = {n_z}\\
 + 2Z_0^{(2)} + {\omega _y} - \omega _1^{(1)} = 0,\;\phi _1^{(1)} = 0,\;n_1^{(1)} = {n_y}\\
 + 2Z_0^{(2)} - {\omega _y} - \omega _2^{(1)} = 0,\;\phi _2^{(1)} = \pi ,\;n_2^{(1)} = {n_y}\\
 + 2Z_0^{(2)} + {\omega _x} - \omega _1^{(2)} = 0,\;\phi _1^{(2)} =  - \pi /2,\;n_1^{(2)} = {n_x}\\
 + 2Z_0^{(2)} - {\omega _x} - \omega _2^{(2)} = 0,\;\phi _2^{(2)} =  + \pi /2,\;n_2^{(2)} = {n_x}\\
{\omega _O} \gg {\omega _X} \ne {\omega _Y} \ne {\omega _Z} \gg 2X_0^{(1)},\;2Z_0^{(2)},\;\omega _i^{(l)}
\end{array}\]
\end{small}
The shared junctions' parameters are shown as:
\begin{small}
${\lambda _O} = {(2{E_{CO}}/{E_{JO}})^{1/4}}$, $\hbar {\omega _O} = {(8{E_{JO}}{E_{CO}})^{1/2}}$, ${E_{CO}} = {e^2}/(2{C_O})$,
${\lambda _P} = {[2{E_{CP}}/({E_{JP}}\cos {\gamma _{P0}} + {E_{Lp}})]^{1/4}} \approx \;{[2{E_{CP}}/{E_{JP}}]^{1/4}}$,
$\hbar {\omega _P} = {[8({E_{JP}}\cos {\gamma _{P0}} + {E_{Lp}}){E_{CP}}]^{1/2}} \approx {[8{E_{JP}}{E_{CP}}]^{1/2}}$,
${E_{CP}} = {e^2}/(2{C_P})$
\end{small}
\par The coupling between momentum and spin in the x direction can be obtained by the iteration where
\\
\begin{small}
${I_1}=  - \hbar X_2^{(1)}\hat \sigma _x^{(1)}({\lambda _X}\hat a_X^ + {e^{ + {\omega _X}it}} + H.C)
({\lambda _O}\hat a_O^ + {e^{ + {\omega _O}it}} + H.C)$
\end{small},
\\
\begin{small}
${I_2}=  + \hbar X_2^{(2)}(\hat \sigma _ + ^{(2)}{e^{ + 2{\varepsilon _2}it}} + H.C)
({\lambda _O}\hat a_O^ + {e^{ + {\omega _O}it}} + H.C)({\textstyle{1 \over 2}}n_i^{(2)}
{e^{ + i(\omega _i^{(2)}t + \phi _i^{(2)})}} + H.C)$
\end{small},
\\
\begin{small}
${I_3}=  - {E_L}({\lambda _x}\hat a_x^ + {e^{ + {\omega _x}it}} + H.C)
({\lambda _X}\hat a_X^ + {e^{ + {\omega _X}it}} + H.C)$
\end{small}.
\begin{small}
\[\begin{array}{l}
{I_1}\int_0^t {{I_2}dt\int_0^t {{I_3}dt} } \\
 =  - \hbar X_2^{(1)}\hat \sigma _x^{(1)}{\lambda _X}\hat a_X^ \pm {e^{ \pm {\omega _X}it}}{\lambda _O}\hat a_O^ \pm {e^{ \pm {\omega _O}it}}\\
 \times \int_0^t {\hbar X_2^{(2)}\hat \sigma _ \pm ^{(2)}{e^{ \pm 2{\varepsilon _2}it}}{\lambda _O}\hat a_O^ \pm {e^{ \pm {\omega _T}it}}{\textstyle{1 \over 2}}n_i^{(2)}{e^{ \pm \omega _i^{(2)}it}}{e^{ \pm \phi _i^{(2)}i}}d{t_1}} \\
 \times \int_0^{{t_1}} {[ - {E_L}{\lambda _x}{\lambda _X}\hat a_x^ \pm \hat a_X^ \pm {e^{( \pm {\omega _x} \pm {\omega _X})it}}]d{t_2}} \\
 = \hbar X_2^{(1)}\hbar X_2^{(2)}{E_L}{\lambda _X}{\lambda _X}{\lambda _O}{\lambda _O}{\lambda _x}{\textstyle{1 \over 2}}n_i^{(2)}\hat a_X^ \pm \hat a_X^ \pm \hat a_O^ \pm \hat a_O^ \pm \hat \sigma _x^{(1)}\hat \sigma _ \pm ^{(2)}\hat a_x^ \pm {e^{ \pm \phi _i^{(2)}i}}\frac{{{e^{( \pm {\omega _X} \pm {\omega _X} \pm {\omega _O} \pm {\omega _O} \pm 2{\varepsilon _2} \pm \omega _i^{(2)} \pm {\omega _x})it}}}}{{( \pm {\omega _O} \pm {\omega _X})i \pm {\omega _X}i}}\\
 \approx \hbar X_2^{(1)}\hbar X_2^{(2)}{E_L}{\lambda _X}{\lambda _X}{\lambda _O}{\lambda _O}{\lambda _x}{\textstyle{1 \over 2}}n_i^{(2)}\hat a_X^ \pm \hat a_X^ \pm \hat a_O^ \pm \hat a_O^ \pm \hat \sigma _x^{(1)}\hat \sigma _ \pm ^{(2)}\hat a_x^ \pm {e^{ \pm \phi _i^{(2)}i}}\frac{{{e^{( \pm {\omega _X} \pm {\omega _X} \pm {\omega _O} \pm {\omega _O} \pm 2{\varepsilon _2} \pm \omega _i^{(2)} \pm {\omega _x})it}}}}{{( \pm {\omega _O})i( \pm {\omega _X})i}}\\
 = \hbar X_2^{(1)}\hbar X_2^{(2)}{E_L}{\lambda _X}{\lambda _X}{\lambda _O}{\lambda _O}{\lambda _x}{\textstyle{1 \over 2}}n_i^{(2)} \times \\
\left( \begin{array}{l}
 + \hat a_X^ - \hat a_X^ + \hat a_O^ - \hat a_O^ + \hat \sigma _x^{(1)}\hat \sigma _ \pm ^{(2)}\hat a_x^ \pm {e^{ \pm \phi _i^{(2)}i}}\frac{{{e^{( \pm 2{\varepsilon _2} \pm \omega _i^{(2)} \pm {\omega _x})it}}}}{{( + {\omega _O})i( + {\omega _X})i}} + \hat a_X^ - \hat a_X^ + \hat a_O^ + \hat a_O^ - \hat \sigma _x^{(1)}\hat \sigma _ \pm ^{(2)}\hat a_x^ \pm {e^{ \pm \phi _i^{(2)}i}}\frac{{{e^{( \pm 2{\varepsilon _2} \pm \omega _i^{(2)} \pm {\omega _x})it}}}}{{( - {\omega _O})i( + {\omega _X})i}}\\
 + \hat a_X^ + \hat a_X^ - \hat a_O^ - \hat a_O^ + \hat \sigma _x^{(1)}\hat \sigma _ \pm ^{(2)}\hat a_x^ \pm {e^{ \pm \phi _i^{(2)}i}}\frac{{{e^{( \pm 2{\varepsilon _2} \pm \omega _i^{(2)} \pm {\omega _x})it}}}}{{( + {\omega _O})i( - {\omega _X})i}} + \hat a_X^ + \hat a_X^ - \hat a_O^ + \hat a_O^ - \hat \sigma _x^{(1)}\hat \sigma _ \pm ^{(2)}\hat a_x^ \pm {e^{ \pm \phi _i^{(2)}i}}\frac{{{e^{( \pm 2{\varepsilon _2} \pm \omega _i^{(2)} \pm {\omega _x})it}}}}{{( - {\omega _O})i( - {\omega _X})i}}
\end{array} \right)\\
 = \hbar X_2^{(1)}\hbar X_2^{(2)}{E_L}\frac{{{\lambda _X}{\lambda _X}{\lambda _O}{\lambda _O}}}{{{\omega _O}i{\omega _X}i}}{\lambda _x}{\textstyle{1 \over 2}}n_i^{(2)}\hat \sigma _x^{(1)}\hat \sigma _ \pm ^{(2)}\hat a_x^ \pm {e^{ \pm \phi _i^{(2)}i}}{e^{( \pm 2{\varepsilon _2} \pm \omega _i^{(2)} \pm {\omega _x})it}}\\
 = \hbar X_2^{(1)}\hbar X_2^{(2)}{E_L}{(4{E_{JX}}{E_{JO}})^{ - 1}}{\lambda _x}{\textstyle{1 \over 2}}n_i^{(2)}\hat \sigma _x^{(1)}\hat \sigma _ \pm ^{(2)}\hat a_x^ \pm {e^{ \pm \phi _i^{(2)}i}}{e^{( \pm 2{\varepsilon _2} \pm \omega _i^{(2)} \pm {\omega _x})it}}
\end{array}\]
\end{small}
\\
\begin{small}
\[\begin{array}{l}
{I_1}\int_0^t {{I_3}dt\int_0^t {{I_2}dt} }  = \\
 - \hbar X_2^{(1)}\hat \sigma _x^{(1)}({\lambda _X}\hat a_X^ \pm {e^{ \pm {\omega _X}it}}{\lambda _O}\hat a_O^ \pm {e^{ \pm {\omega _O}it}})\\
 \times \int_0^t {[ - {E_L}{\lambda _x}{\lambda _X}\hat a_x^ \pm \hat a_X^ \pm {e^{( \pm {\omega _x} \pm {\omega _X})it}}]dt} \\
 \times \int_0^t {\hbar X_2^{(2)}\hat \sigma _ \pm ^{(2)}{e^{ \pm 2{\varepsilon _2}it}}({\lambda _O}\hat a_O^ \pm {e^{ \pm {\omega _O}it}}{\textstyle{1 \over 2}}n_i^{(2)}{e^{ \pm \omega _i^{(2)}it}}{e^{ \pm \phi _i^{(2)}i}})dt} \\
 = \hbar X_2^{(1)}\hbar X_2^{(2)}{E_L}{\lambda _X}{\lambda _X}{\lambda _O}{\lambda _O}{\lambda _x}{\textstyle{1 \over 2}}n_i^{(2)}\hat a_X^ \pm \hat a_X^ \pm \hat a_O^ \pm \hat a_O^ \pm \hat \sigma _x^{(1)}\hat \sigma _ \pm ^{(2)}\hat a_x^ \pm {e^{ \pm \phi _i^{(2)}i}}\frac{{{e^{( \pm {\omega _X} \pm {\omega _X} \pm {\omega _O} \pm {\omega _O} \pm {\omega _x} \pm \omega _i^{(2)} \pm 2{\varepsilon _2})it}}}}{{( \pm {\omega _X} \pm {\omega _O})i( \pm {\omega _O})i}}\\
 \approx \hbar X_2^{(1)}\hbar X_2^{(2)}{E_L}{\lambda _X}{\lambda _X}{\lambda _O}{\lambda _O}{\lambda _x}{\textstyle{1 \over 2}}n_i^{(2)}\hat a_X^ \pm \hat a_X^ \pm \hat a_O^ \pm \hat a_O^ \pm \hat \sigma _x^{(1)}\hat \sigma _ \pm ^{(2)}\hat a_x^ \pm {e^{ \pm \phi _i^{(2)}i}}\frac{{{e^{( \pm {\omega _X} \pm {\omega _X} \pm {\omega _O} \pm {\omega _O} \pm {\omega _x} \pm \omega _i^{(2)} \pm 2{\varepsilon _2})it}}}}{{( \pm {\omega _O})i( \pm {\omega _O})i}}
\end{array}\]
\end{small}
\\
\begin{small}
\[\begin{array}{l}
{I_2}\int_0^t {{I_1}dt\int_0^t {{I_3}dt} }  = \\
 = X_2^{(2)}\hat \sigma _ \pm ^{(2)}{e^{ \pm 2{\varepsilon _2}it}}{\lambda _O}\hat a_O^ \pm {e^{ \pm {\omega _T}it}}{\textstyle{1 \over 2}}n_i^{(2)}{e^{ \pm \omega _i^{(2)}it}}{e^{ \pm \phi _i^{(2)}i}}\\
 \times \int_0^t {[ - X_2^{(1)}\hat \sigma _x^{(1)}{\lambda _X}\hat a_X^ \pm {e^{ \pm {\omega _X}it}}{\lambda _O}\hat a_O^ \pm {e^{ \pm {\omega _O}it}}]d{t_1}} \\
 \times \int_0^{{t_1}} {[ - {E_L}{\lambda _x}{\lambda _X}\hat a_x^ \pm \hat a_X^ \pm {e^{( \pm {\omega _x} \pm {\omega _X})it}}]d{t_2}} \\
 = \hbar X_2^{(2)}\hbar X_2^{(1)}{E_L}{\lambda _X}{\lambda _X}{\lambda _O}{\lambda _O}{\lambda _x}{\textstyle{1 \over 2}}n_i^{(2)}\hat a_X^ \pm \hat a_X^ \pm \hat a_O^ \pm \hat a_O^ \pm \hat \sigma _x^{(1)}\hat \sigma _ \pm ^{(2)}\hat a_x^ \pm {e^{ \pm \phi _i^{(2)}i}}\frac{{{e^{( \pm {\omega _X} \pm {\omega _X} \pm {\omega _O} \pm {\omega _O} \pm {\omega _x} \pm \omega _i^{(2)} \pm 2{\varepsilon _2})it}}}}{{( \pm {\omega _X} \pm {\omega _O} \pm {\omega _X})i( \pm {\omega _X})i}}\\
 = \hbar X_2^{(2)}\hbar X_2^{(1)}{E_L}{\lambda _X}{\lambda _X}{\lambda _O}{\lambda _O}{\lambda _x}{\textstyle{1 \over 2}}n_i^{(2)} \times \\
\left( \begin{array}{l}
 + \hat a_X^ - \hat a_X^ + \hat a_O^ - \hat a_O^ + \hat \sigma _x^{(1)}\hat \sigma _ \pm ^{(2)}\hat a_x^ \pm {e^{ \pm \phi _i^{(2)}i}}\frac{{{e^{( \pm {\omega _x} \pm \omega _i^{(2)} \pm 2{\varepsilon _2})it}}}}{{( + {\omega _O})i( + {\omega _X})i}} + \hat a_X^ - \hat a_X^ + \hat a_O^ + \hat a_O^ - \hat \sigma _x^{(1)}\hat \sigma _ \pm ^{(2)}\hat a_x^ \pm {e^{ \pm \phi _i^{(2)}i}}\frac{{{e^{( \pm {\omega _x} \pm \omega _i^{(2)} \pm 2{\varepsilon _2})it}}}}{{( - {\omega _O})i( + {\omega _X})i}}\\
 + \hat a_X^ + \hat a_X^ - \hat a_O^ - \hat a_O^ + \hat \sigma _x^{(1)}\hat \sigma _ \pm ^{(2)}\hat a_x^ \pm {e^{ \pm \phi _i^{(2)}i}}\frac{{{e^{( \pm {\omega _x} \pm \omega _i^{(2)} \pm 2{\varepsilon _2})it}}}}{{( + {\omega _O})i( - {\omega _X})i}} + \hat a_X^ + \hat a_X^ - \hat a_O^ + \hat a_O^ - \hat \sigma _x^{(1)}\hat \sigma _ \pm ^{(2)}\hat a_x^ \pm {e^{ \pm \phi _i^{(2)}i}}\frac{{{e^{( \pm {\omega _x} \pm \omega _i^{(2)} \pm 2{\varepsilon _2})it}}}}{{( - {\omega _O})i( - {\omega _X})i}}
\end{array} \right)\\
 = \hbar X_2^{(2)}\hbar X_2^{(1)}{E_L}\frac{{{\lambda _X}{\lambda _X}{\lambda _O}{\lambda _O}}}{{{\omega _O}i{\omega _X}i}}{\lambda _x}{\textstyle{1 \over 2}}n_i^{(2)}\hat \sigma _x^{(1)}\hat \sigma _ \pm ^{(2)}\hat a_x^ \pm {e^{ \pm \phi _i^{(2)}i}}{e^{( \pm {\omega _x} \pm \omega _i^{(2)} \pm 2{\varepsilon _2})it}}\\
 = \hbar X_2^{(2)}\hbar X_2^{(1)}{E_L}{(4{E_{JX}}{E_{JO}})^{ - 1}}{\lambda _x}{\textstyle{1 \over 2}}n_i^{(2)}\hat \sigma _x^{(1)}\hat \sigma _ \pm ^{(2)}\hat a_x^ \pm {e^{ \pm \phi _i^{(2)}i}}{e^{( \pm {\omega _x} \pm \omega _i^{(2)} \pm 2{\varepsilon _2})it}}
\end{array}\]
\end{small}
\\
\begin{small}
\[\begin{array}{l}
{I_3}\int_0^t {{I_1}dt\int_0^t {{I_2}dt} }  = \\
 =  - {E_L}{\lambda _x}{\lambda _X}\hat a_x^ \pm \hat a_X^ \pm {e^{( \pm {\omega _x} \pm {\omega _X})it}}\\
 \times \int_0^t {[ - X_2^{(1)}\hat \sigma _x^{(1)}{\lambda _X}\hat a_X^ \pm {e^{ \pm {\omega _X}it}}{\lambda _O}\hat a_O^ \pm {e^{ \pm {\omega _O}it}}]d{t_1}} \\
 \times \int_0^{{t_1}} {X_2^{(2)}\hat \sigma _ \pm ^{(2)}{e^{ \pm 2{\varepsilon _2}it}}{\lambda _O}\hat a_O^ \pm {e^{ \pm {\omega _O}it}}{\textstyle{1 \over 2}}n_i^{(2)}{e^{ \pm \omega _i^{(2)}it}}{e^{ \pm \phi _i^{(2)}i}}d{t_2}} \\
 = \hbar X_2^{(1)}\hbar X_2^{(2)}{E_L}{\lambda _X}{\lambda _X}{\lambda _O}{\lambda _O}{\lambda _x}{\textstyle{1 \over 2}}n_i^{(2)}\hat a_X^ \pm \hat a_X^ \pm \hat a_O^ \pm \hat a_O^ \pm \hat \sigma _x^{(1)}\hat \sigma _ \pm ^{(2)}\hat a_x^ \pm {e^{ \pm \phi _i^{(2)}i}}\frac{{{e^{( \pm {\omega _O} \pm {\omega _O} \pm {\omega _X} \pm {\omega _X} \pm {\omega _x} \pm 2{\varepsilon _2} \pm \omega _i^{(2)})i{t_2}}}}}{{( \pm {\omega _X} \pm {\omega _O} \pm {\omega _O})i( \pm {\omega _O})i}}\\
 = \hbar X_2^{(1)}\hbar X_2^{(2)}{E_L}{\lambda _X}{\lambda _X}{\lambda _O}{\lambda _O}{\lambda _x}{\textstyle{1 \over 2}}n_i^{(2)} \times \\
\left( \begin{array}{l}
 + \hat a_X^ - \hat a_X^ + \hat a_O^ - \hat a_O^ + \hat \sigma _x^{(1)}\hat \sigma _ \pm ^{(2)}\hat a_x^ \pm {e^{ \pm \phi _i^{(2)}i}}\frac{{{e^{( \pm {\omega _x} \pm 2{\varepsilon _2} \pm \omega _i^{(2)})i{t_2}}}}}{{( + {\omega _X})i( + {\omega _O})i}} + \hat a_X^ + \hat a_X^ - \hat a_O^ - \hat a_O^ + \hat \sigma _x^{(1)}\hat \sigma _ \pm ^{(2)}\hat a_x^ \pm {e^{ \pm \phi _i^{(2)}i}}\frac{{{e^{( \pm {\omega _x} \pm 2{\varepsilon _2} \pm \omega _i^{(2)})i{t_2}}}}}{{( - {\omega _X})i( + {\omega _O})i}}\\
 + \hat a_X^ - \hat a_X^ + \hat a_O^ + \hat a_O^ - \hat \sigma _x^{(1)}\hat \sigma _ \pm ^{(2)}\hat a_x^ \pm {e^{ \pm \phi _i^{(2)}i}}\frac{{{e^{( \pm {\omega _x} \pm 2{\varepsilon _2} \pm \omega _i^{(2)})i{t_2}}}}}{{( + {\omega _X})i( - {\omega _O})i}} + \hat a_X^ + \hat a_X^ - \hat a_O^ + \hat a_O^ - \hat \sigma _x^{(1)}\hat \sigma _ \pm ^{(2)}\hat a_x^ \pm {e^{ \pm \phi _i^{(2)}i}}\frac{{{e^{( \pm {\omega _x} \pm 2{\varepsilon _2} \pm \omega _i^{(2)})i{t_2}}}}}{{( - {\omega _X})i( - {\omega _O})i}}
\end{array} \right)\\
 = \hbar X_2^{(1)}\hbar X_2^{(2)}{E_L}\frac{{{\lambda _X}{\lambda _X}{\lambda _O}{\lambda _O}}}{{( + {\omega _X})i( + {\omega _O})i}}{\lambda _x}{\textstyle{1 \over 2}}n_i^{(2)}\hat \sigma _x^{(1)}\hat \sigma _ \pm ^{(2)}\hat a_x^ \pm {e^{ \pm \phi _i^{(2)}i}}{e^{( \pm {\omega _x} \pm 2{\varepsilon _2} \pm \omega _i^{(2)})i{t_2}}}\\
 = \hbar X_2^{(1)}\hbar X_2^{(2)}{E_L}{(4{E_{JX}}{E_{JO}})^{ - 1}}{\lambda _x}{\textstyle{1 \over 2}}n_i^{(2)}\hat \sigma _x^{(1)}\hat \sigma _ \pm ^{(2)}\hat a_x^ \pm {e^{ \pm \phi _i^{(2)}i}}{e^{( \pm {\omega _x} \pm 2{\varepsilon _2} \pm \omega _i^{(2)})i{t_2}}}
\end{array}\]
\end{small}
\\
\begin{small}
\[\begin{array}{l}
{I_2}\int_0^t {{I_3}dt\int_0^t {{I_1}dt} }  = \\
 = X_2^{(2)}\hat \sigma _ \pm ^{(2)}{e^{ \pm 2{\varepsilon _2}it}}{\lambda _O}\hat a_O^ \pm {e^{ \pm {\omega _T}it}}{\textstyle{1 \over 2}}n_i^{(2)}{e^{ \pm \omega _i^{(2)}it}}{e^{ \pm \phi _i^{(2)}i}}\\
 \times \int_0^t {[ - {E_L}{\lambda _x}{\lambda _X}\hat a_x^ \pm \hat a_X^ \pm {e^{( \pm {\omega _x} \pm {\omega _X})i{t_1}}}]d{t_1}} \\
 \times \int_0^{{t_1}} {[ - X_2^{(1)}\hat \sigma _x^{(1)}{\lambda _X}{\lambda _O}\hat a_X^ \pm \hat a_O^ \pm {e^{( \pm {\omega _X} \pm {\omega _O})i{t_2}}}]d{t_2}} \\
 = \hbar X_2^{(1)}\hbar X_2^{(2)}{E_L}{\lambda _X}{\lambda _X}{\lambda _O}{\lambda _O}{\lambda _x}{\textstyle{1 \over 2}}n_i^{(2)}\hat a_X^ \pm \hat a_X^ \pm \hat a_O^ \pm \hat a_O^ \pm \hat \sigma _x^{(1)}\hat \sigma _ \pm ^{(2)}\hat a_x^ \pm {e^{ \pm \phi _i^{(2)}i}}\frac{{{e^{( \pm {\omega _O} \pm {\omega _O} \pm {\omega _X} \pm {\omega _X} \pm 2{\varepsilon _2} \pm {\omega _x} \pm \omega _i^{(2)})i{t_1}}}}}{{( \pm {\omega _X} \pm {\omega _X} \pm {\omega _O})i( \pm {\omega _X} \pm {\omega _O})i}}\\
 \approx \hbar X_2^{(1)}\hbar X_2^{(2)}{E_L}{\lambda _X}{\lambda _X}{\lambda _O}{\lambda _O}{\lambda _x}{\textstyle{1 \over 2}}n_i^{(2)}\hat a_X^ \pm \hat a_X^ \pm \hat a_O^ \pm \hat a_O^ \pm \hat \sigma _x^{(1)}\hat \sigma _ \pm ^{(2)}\hat a_x^ \pm {e^{ \pm \phi _i^{(2)}i}}\frac{{{e^{( \pm {\omega _T} \pm {\omega _T} \pm {\omega _X} \pm {\omega _X} \pm 2{\varepsilon _2} \pm {\omega _x} \pm \omega _i^{(2)})i{t_1}}}}}{{( \pm {\omega _O})i( \pm {\omega _O})i}}
\end{array}\]
\end{small}
\\
\begin{small}
\[\begin{array}{l}
{I_3}\int_0^t {{I_2}dt\int_0^t {{I_1}dt} }  = \\
 =  - {E_L}{\lambda _x}{\lambda _X}\hat a_x^ \pm \hat a_X^ \pm {e^{( \pm {\omega _x} \pm {\omega _X})it}}\\
 \times \int_0^t {X_2^{(2)}\hat \sigma _ \pm ^{(2)}{e^{ \pm 2{\varepsilon _2}it}}{\lambda _O}\hat a_O^ \pm {e^{ \pm {\omega _O}it}}{\textstyle{1 \over 2}}n_i^{(2)}{e^{ \pm \omega _i^{(2)}it}}{e^{ \pm \phi _i^{(2)}i}}d{t_1}} \\
 \times \int_0^{{t_1}} {[ - X_2^{(1)}\hat \sigma _x^{(1)}{\lambda _X}\hat a_X^ \pm {e^{ \pm {\omega _X}it}}{\lambda _O}\hat a_O^ \pm {e^{ \pm {\omega _T}it}}]d{t_2}} \\
 = \hbar X_2^{(2)}\hbar X_2^{(1)}{E_L}{\lambda _X}{\lambda _X}{\lambda _O}{\lambda _O}{\lambda _x}{\textstyle{1 \over 2}}n_i^{(2)}\hat a_X^ \pm \hat a_X^ \pm \hat a_O^ \pm \hat a_O^ \pm \hat \sigma _x^{(1)}\hat \sigma _ \pm ^{(2)}\hat a_x^ \pm {e^{ \pm \phi _i^{(2)}i}}\frac{{{e^{( \pm {\omega _X} \pm {\omega _X} \pm {\omega _O} \pm {\omega _O} \pm {\omega _x} \pm 2{\varepsilon _2} \pm \omega _i^{(2)})i{t_2}}}}}{{( \pm {\omega _O} \pm {\omega _X} \pm {\omega _O})i( \pm {\omega _X} \pm {\omega _O})i}}\\
 \approx \hbar X_2^{(2)}\hbar X_2^{(1)}{E_L}{\lambda _X}{\lambda _X}{\lambda _O}{\lambda _O}{\lambda _x}{\textstyle{1 \over 2}}n_i^{(2)}\hat a_X^ \pm \hat a_X^ \pm \hat a_O^ \pm \hat a_O^ \pm \hat \sigma _x^{(1)}\hat \sigma _ \pm ^{(2)}\hat a_x^ \pm {e^{ \pm \phi _i^{(2)}i}}\frac{{{e^{( \pm {\omega _X} \pm {\omega _X} \pm {\omega _T} \pm {\omega _T} \pm {\omega _x} \pm 2{\varepsilon _2} \pm \omega _i^{(2)})i{t_2}}}}}{{( \pm {\omega _O} \pm {\omega _O} \pm {\omega _X})i( \pm {\omega _O})i}}\\
 = \hbar X_2^{(2)}\hbar X_2^{(1)}{E_L}{\lambda _X}{\lambda _X}{\lambda _O}{\lambda _O}{\lambda _x}{\textstyle{1 \over 2}}n_i^{(2)} \times \\
\left( \begin{array}{l}
 + \hat a_X^ - \hat a_X^ + \hat a_O^ - \hat a_O^ + \hat \sigma _x^{(1)}\hat \sigma _ \pm ^{(2)}\hat a_x^ \pm {e^{ \pm \phi _i^{(2)}i}}\frac{{{e^{( \pm {\omega _x} \pm 2{\varepsilon _2} \pm \omega _i^{(2)})i{t_2}}}}}{{( + {\omega _X})i( + {\omega _O})i}} + \hat a_X^ + \hat a_X^ - \hat a_O^ - \hat a_O^ + \hat \sigma _x^{(1)}\hat \sigma _ \pm ^{(2)}\hat a_x^ \pm {e^{ \pm \phi _i^{(2)}i}}\frac{{{e^{( \pm {\omega _x} \pm 2{\varepsilon _2} \pm \omega _i^{(2)})i{t_2}}}}}{{( - {\omega _X})i( + {\omega _O})i}}\\
 + \hat a_X^ - \hat a_X^ + \hat a_O^ + \hat a_O^ - \hat \sigma _x^{(1)}\hat \sigma _ \pm ^{(2)}\hat a_x^ \pm {e^{ \pm \phi _i^{(2)}i}}\frac{{{e^{( \pm {\omega _x} \pm 2{\varepsilon _2} \pm \omega _i^{(2)})i{t_2}}}}}{{( + {\omega _X})i( - {\omega _O})i}} + \hat a_X^ + \hat a_X^ - \hat a_O^ + \hat a_O^ - \hat \sigma _x^{(1)}\hat \sigma _ \pm ^{(2)}\hat a_x^ \pm {e^{ \pm \phi _i^{(2)}i}}\frac{{{e^{( \pm {\omega _x} \pm 2{\varepsilon _2} \pm \omega _i^{(2)})i{t_2}}}}}{{( - {\omega _X})i( - {\omega _O})i}}
\end{array} \right)\\
 = \hbar X_2^{(2)}\hbar X_2^{(1)}{E_L}\frac{{{\lambda _X}{\lambda _X}{\lambda _O}{\lambda _O}}}{{( + {\omega _X})i( + {\omega _O})i}}{\lambda _x}{\textstyle{1 \over 2}}n_i^{(2)}\hat \sigma _x^{(1)}\hat \sigma _ \pm ^{(2)}\hat a_x^ \pm {e^{ \pm \phi _i^{(2)}i}}{e^{( \pm {\omega _x} \pm 2{\varepsilon _2} \pm \omega _i^{(2)})i{t_2}}}\\
 = \hbar X_2^{(2)}\hbar X_2^{(1)}{E_L}{(4{E_{JX}}{E_{JO}})^{ - 1}}{\lambda _x}{\textstyle{1 \over 2}}n_i^{(2)}\hat \sigma _x^{(1)}\hat \sigma _ \pm ^{(2)}\hat a_x^ \pm {e^{ \pm \phi _i^{(2)}i}}{e^{( \pm {\omega _x} \pm 2{\varepsilon _2} \pm \omega _i^{(2)})i{t_2}}}
\end{array}\]
\end{small}
The coupling terms \begin{small}${I_1}\int_0^t {{I_3}dt\int_0^t {{I_2}dt} }$\end{small} and
\begin{small}${I_2}\int_0^t {{I_3}dt\int_0^t {{I_1}dt} }$\end{small} are discarded. The pulses with
\begin{small}$\omega _2^{(2)}$, $\omega _1^{(2)}$\end{small} excite the resonant JC coupling and AJC coupling
respectively. So, the coupling between momentum and spin in the x direction is shown as
\begin{small}
${H_{I3x}} = {\textstyle{1 \over 2}}\hbar X_2^{(1)}\hbar X_2^{(2)}{E_L}{({E_{JO}}{E_{JX}})^{ - 1}}
{\lambda _x}{n_x}\hat \sigma _x^{(1)}\hat \sigma _x^{(2)}i(\hat a_x^ +  - \hat a_x^ - )$
\end{small}
\par The coupling between momentum and spin in the y direction can be obtained by the iteration where
\\
\begin{small}
${I_1} =  - \hbar X_2^{(1)}\hat \sigma _x^{(1)}({\lambda _O}\hat a_O^ + {e^{ + {\omega _O}it}} + H.C)
({\textstyle{1 \over 2}}n_i^{(1)}{e^{ + i(\omega _i^{(1)}t + \phi _i^{(1)})}} + H.C)$
\end{small},
\\
\begin{small}
${I_2} =  + \hbar X_2^{(2)}(\hat \sigma _ + ^{(2)}{e^{ + 2{\varepsilon _2}it}} + H.C)
({\lambda _Y}\hat a_Y^ + {e^{ + {\omega _Y}it}} + H.C)({\lambda _O}\hat a_O^ + {e^{ + {\omega _O}it}} + H.C)$
\end{small},
\\
\begin{small}
${I_3} =  - {E_L}({\lambda _y}\hat a_y^ + {e^{ + {\omega _y}it}} + H.C)
({\lambda _Y}\hat a_Y^ + {e^{ + {\omega _Y}it}} + H.C)$
\end{small}.
\\
The pulses with
\begin{small}$\omega _1^{(1)}$, $\omega _2^{(1)}$\end{small} excite the resonant JC coupling and AJC coupling
respectively. So, the coupling between momentum and spin in the y direction is shown as
\begin{small}
${H_{I3y}} = {\textstyle{1 \over 2}}\hbar X_2^{(1)}\hbar X_2^{(2)}{E_L}{({E_{JO}}{E_{JY}})^{ - 1}}
{\lambda _y}{n_y}\hat \sigma _x^{(1)}\hat \sigma _y^{(2)}i(\hat a_y^ +  - \hat a_y^ - )$
\end{small}
\par The coupling between momentum and spin in the z direction can be obtained by the iteration where
\\
\begin{small}
${I_1} =  - \hbar X_2^{(1)}\hat \sigma _x^{(1)}({\lambda _Z}\hat a_Z^ + {e^{ + {\omega _Z}it}} + H.C)
({\lambda _O}\hat a_O^ + {e^{ + {\omega _O}it}} + H.C)$
\end{small},
\\
\begin{small}
${I_2} =  + \hbar Z_2^{(2)}\hat \sigma _z^{(2)}({\lambda _O}\hat a_O^ + {e^{ + {\omega _O}it}} + H.C)
({\textstyle{1 \over 2}}n_i^{(2)}{e^{ + i(\omega _i^{(2)}t + \phi _i^{(2)})}} + H.C)$
\end{small},
\\
\begin{small}
${I_3} =  - {E_L}({\lambda _z}\hat a_z^ + {e^{ + {\omega _z}it}} + H.C)
({\lambda _Z}\hat a_Z^ + {e^{ + {\omega _Z}it}} + H.C)$
\end{small}.
\\
The pulse with
\begin{small}$\omega _3^{(2)}$\end{small} excite the resonant JC coupling and AJC coupling
respectively. So, the coupling between momentum and spin in the z direction is shown as
\begin{small}
${H_{I3Z}} = {\textstyle{1 \over 2}}\hbar X_2^{(1)}\hbar Z_2^{(2)}{E_L}{({E_{JO}}{E_{JZ}})^{ - 1}}
{\lambda _z}{n_z}{\kern 1pt} \hat \sigma _x^{(1)}\hat \sigma _z^{(2)}i(\hat a_z^ +  - \hat a_z^ - )$
\end{small}.
\par In 2+1D, all items are same as that in 3+1D, except \begin{small}${H_{I3Z}} =0$\end{small}.
In 1+1D, the first order approximation is the same as that in 3+1D.
The second order approximation represents the resonant coupling term
between the flux qubit and phase qubit.
It can be obtained by iteration of
\begin{small}${H_{I2}} = \sum {{I_k}\int_0^{{t_1}} {{I_m}d{t_2}} }$\end{small},
where \begin{small}$k,m=1,2$\end{small} and \begin{small}$k \ne m$\end{small}.
\\
\begin{small}
${I_1} = \hbar {X_2}{{\hat \sigma }_x}({\lambda _X}\hat a_X^ + {e^{ + {\omega _X}it}} + H.C)
({\textstyle{1 \over 2}}{n_i}{e^{i({\omega _i}t + {\phi _i})}} + H.C)$
\end{small},\\
\begin{small}
${I_2} =- {E_L}({\lambda _x}\hat a_x^ + {e^{ + {\omega _x}it}} + H.C)
({\lambda _X}\hat a_X^ + {e^{ + {\omega _X}it}} + H.C)$
\end{small}.\\
It is shown as
\begin{small}
${H_{I2}}={\textstyle{1 \over 2}}\hbar X_2^{(1)}{E_L}E_{JX}^{( - 1)}
{\lambda _x}{n_1}{{\hat \sigma }_x}i({{\hat a}_x} - \hat a_x^ + )$
\end{small}.

\begin{thebibliography}{99}
\bibitem{1}L. Lamata, J. Le¨®n, T. Sch$\ddot a$tz, and E. Solano,
Physical Review Letters {\bf 98}, 253005(2007).
\bibitem{2}A. Bermudez, M. A. Martin-Delgado, and E. Solano,
Physical Review A {\bf 76}, 041801(R)(2007).
\bibitem{3}R. Gerritsma1,2, G. Kirchmair, F.Z$\ddot a$hringer, E. Solano, R. Blatt and C.
F. Roos, nature {\bf 463}, 68(2010).
\bibitem{4}Xiangdong Zhang, Physical Review Letters {\bf 100}, 113903(2008).
\bibitem{5}Xiangdong Zhang and Zhengyou Liu, Physical Review Letters {\bf 101}, 264303(2008).
\bibitem{10}I. Chiorescu, P. Bertet, K. Semba, Y. Nakamura,
C. J. P. M. Harmans, J. E. Mooij, Nature {\bf 431}, 159(2004).
\bibitem{11}J. Q. You, Franco Nori, Nature {\bf 474}, 589(2011).
\bibitem{12}A. Bermudez, M. A. Martin-Delgado, and E. Solano,
Physical Review Letters {\bf 99}, 123602(2007).
\bibitem{13}J¨®zsef Cserti and Gyula D¨¢vid,
Physical Review B {\bf 74}, 172305(2006).
\bibitem{15}Yu-Xi Liu, J. Q. You, L. F. Wei, C. P. Sun, and Franco. Nori,
Physical Review Letters {\bf 95}, 087001(2005).
\bibitem{16}Yuriy Makhlin, Gerd Schon, Alexander Shnirman,
Reviews of Modern Physics {\bf 73}, 357(2001).
\bibitem{17}Iulia Buluta, Franco Nori, Science {\bf 326}, 108(2009).
I. M. Georgescu, S. Ashhab, Franco Nori,
Reviews of Modern Physics  {\bf 86}, 153.
\bibitem{19}Abhinav Kandala, Antonio Mezzacapo, Kristan Temme, Maika
Takita, Markus Brink, Jerry M. Chow, Jay M. Gambetta,
Nature{\bf 549}, 242(2017).
\bibitem{20}J. E. Mooij, T. P. Orlando, L. Levitov, Lin Tian,
Caspar H. van der Wal, Seth Lloyd,
Science{\bf 285}, 1036(1999).
\bibitem{21}T. P. Orlando, J. E. Mooij, Lin Tian, Caspar H. van der Wal,
L. S. Levitov, Seth Lloyd, J. J. Mazo,
Physical Review B {\bf 60}, 15398(1999).
\bibitem{22}A. O. Niskanen, K. Harrabi, F. Yoshihara, Y. Nakamura, and J. S. Tsai,
Physical Review B {\bf 74}, 220503(R)(2006).
\bibitem{23}J. Q. You, Xuedong Hu, Ashhab, and Franco Nori,
Physical Review B {\bf 75}, 140515(R)(2007).
\bibitem{25}Xiao-Ling He, J. Q. You, Yu-xi Liu, L. F. Wei and Franco Nori,
Physical Review B {\bf 76}, 024517(2007).
\bibitem{26}Yu-xi Liu, L. F. Wei, J. S. Tsai, and Franco Nori,
Physical Review Letters {\bf 96}, 067003(2006).
\bibitem{27}J. Q. You, Y. Nakamura, and Franco Nori,
Physical Review B {\bf 71}, 024532(2005).
\bibitem{29}Kaushik Mitra, F. W. Strauch, C. J. Lobb, J. R. Anderson, and F. C. Wellstood,
Physical Review B {\bf 77}, 214512(2008).
\bibitem{31}Alec Maassen van den Brink,
Physical Review B {\bf 71}, 064503(2005).
\bibitem{33}Samuele Spilla, Fabian Hassler and Janine Splettstoesser,
New Journal of Physics {\bf 16}, 045020(2014).
\bibitem{35}E. Paladino, Y. M. Galperin, G. Falci, B. L. Altshuler,
Reviews of Modern Physics {\bf 86}, 361(2014).
\bibitem{36}S. Berger, M. Pechal, P. Kurpiers, A.A. Abdumalikov,
C. Eichler,w, J.A. Mlynek, A. Shnirman, Yuval Gefen, A. Wallraff, S. Filipp,
Nature Communications {\bf 6}, 8757
\bibitem{37}P.J.Leek, J. M. Fink, A. Blais, R. Bianchetti, M.G$\ddot o$ppl, J.M.Gambetta,
D.I.Schuster, L.Frunzio, R.J.Schoelkopf, A.Wallraff
Science {\bf 318}, 1889(2007).
\bibitem{38}U. Patel, Y. Gao, D. Hover, G. J. Ribeill, S. Sendelbach, and R. McDermott,
Applied Physics Letters {\bf 102}, 012602(2013).
\bibitem{39}J. Q. You, J. S. Tsai, and Franco Nori,
Physical Review B {\bf 73}, 014510(2006).
\bibitem{41}P. Bertet, I. Chiorescu, G. Burkard, K. Semba,
C. J. P. M. Harmans, D. P. DiVincenzo, and J. E. Mooij,
Physical Review Letters {\bf 95}, 257002(2005).
\bibitem{47}Jonas Bylander, Simon Gustavsson, Fei Yan, Fumiki Yoshihara, Khalil Harrabi, George Fitch,
David G. Cory Yasunobu Nakamura, Jaw-Shen Tsai and William D. Oliver,
Nature Physics {\bf 7}, 565(2011).
\bibitem{49}J. K. Julin, P. J. Koppinen, and I. J. Maasilta,
Applied Physics Letters {\bf 97}, 152501(2010);
\bibitem{51}A. Lupascu, E. F. C. Driessen, L. Roschier, C. J. P. M. Harmans, and J. E. Mooij,
Physical Review Letters {\bf 96}, 127003(2006).
\bibitem{52}Zhongyuan Zhou, Shih-I Chu and Siyuan Han, J.
Phys. B: At. Mol. Opt. Phys. {\bf 41}, 045506(2008).
\end{thebibliography}
\end{document}